\documentclass[prd,eqsecnum,floats,nopacs]{revtex4}
\usepackage{amsmath}
\usepackage{amssymb}
\usepackage{amsfonts}
\usepackage{dsfont}
\usepackage{graphicx, epsfig}
\DeclareMathAlphabet{\mathpzc}{OT1}{pzc}{m}{it}

\newcommand{\bs}[2]{{{\sf{b}}
\phantom{]}}^{\!\!\!\!\!\!*{\kern 1.3pt}{\mbox{${\scriptstyle #1}$}}}_{\mbox{${\scriptstyle #2}$}}}
\newcommand{\as}[2]{{{\sf{a}}
\phantom{]}}^{\!\!\!\!\!\!*{\kern 1.3pt}{\mbox{${\scriptstyle #1}$}}}_{\mbox{${\scriptstyle #2}$}}}
\newcommand{\naf}[2]{n'^{\!\!\!\!\!*{\kern 2.8pt}{\mbox{${\scriptstyle(#1)}$}}}
_{\mbox{${\scriptstyle #2}$}}}
\newcommand{\na}[2]{n^{\!\!\!\!*{\kern 1.3pt}{\mbox{${\scriptstyle(#1)}$}}}
_{\mbox{${\scriptstyle #2}$}}}
\renewcommand{\ap}[2]{a^{\!\!\!\!*{\kern 1.3pt}{\mbox{${\scriptstyle(#1)}$}}}
_{\mbox{${\scriptstyle #2}$}}}
\newcommand{\ep}[1]{e^{\!\!\!\!*{\kern 1.3pt}{\mbox{${\phantom{()}}$}}}
_{\mbox{${\scriptstyle #1}$}}}

\newcommand{\ad}[2]{a^{\!\!\!\!*{\kern 1.3pt}{\mbox{${\scriptstyle #1}$}}}_{\mbox{${\scriptstyle #2}$}}}

\newcommand{\gug}[3]{{#1}^{\!\!\!\!*{\kern 1.3pt}{\mbox{${\scriptstyle #2}$}}}_{\mbox{${\scriptstyle #3}$}}}

\newcommand{\const}{\mathrm{const}}

\newcommand{\Tr}{\mathrm{Tr}}

\newcommand {\oks}[2]{{\raise0.7ex\hbox{${\scriptstyle #1}$}\!\mathord{\left/
{\vphantom{{1}{2}}}\right.\kern-\nulldelimiterspace}
\!\lower0.7ex
\hbox{${\scriptstyle #2}$}}}
\newcommand {\bb}[1]{\mbox{\boldmath $#1$}}
\newcommand{\ii}{\mathrm{i}}

\begin{document}

\title{Neutrino flavor oscillations and spin rotation in matter and electromagnetic field}
\author{A. V. Chukhnova}\email{av.chukhnova@physics.msu.ru}
\author{A. E. Lobanov}\email{lobanov@phys.msu.ru}

\affiliation {Department of Theoretical Physics, Faculty of Physics,
  Moscow State University, 119991 Moscow, Russia}

\begin{abstract}
We obtain a relativistically covariant wave equation for neutrinos in dense matter and electromagnetic field, which describes both flavor oscillations and neutrino spin rotation. Using this equation we construct a quasi-classical theory of these phenomena. We obtain the probabilities of arbitrary spin-flavor transitions assuming the external conditions to be constant. We demonstrate that the resonance behavior of the transition probabilities is possible only when the neutrino flavor states cannot be described as superpositions of the mass eigenstates. We discover that a resonance, which is similar to the Mikheev--Smirnov--Wolfenstein resonance, takes place for neutrinos in magnetic field due to the transition magnetic moments. This resonance gives an opportunity to determine, whether neutrinos are Dirac or Majorana particles.
\end{abstract}

\maketitle

\section{INTRODUCTION}

The phenomenological theory of neutrino oscillations in vacuum is based on the pioneer works by
B. Pontecorvo (see \cite{Bilenky1978}). The study of the influence of external conditions is the next important step toward understanding the nature of neutrino.
The first significant result in this field was obtained by Wolfenstein \cite{Wolfenstein1978}. In his paper the neutrino interaction with the medium was described as a collective effect. Such interaction is associated with the forward elastic scattering of neutrinos by the fermions of the medium. It is described by an effective potential, which modifies the dispersion law. On the base of this approach the Mikheev--Smirnov--Wolfenstein (MSW) effect \cite{MS_en} was discovered.

However, to describe the neutrino evolution it is also important to take into account the electromagnetic properties of neutrinos. Being a massive neutral particle, a Dirac neutrino is characterized by an anomalous magnetic moment. In addition to the diagonal anomalous magnetic moments \cite{Fujikawa1980} neutrinos are characterized by the so-called transition (non-diagonal) moments \cite{Shrock1982} (see also \cite{Giunti2015} and references therein), which affect the mixing of different neutrino mass eigenstates.

One of the reasons of neutrino spin rotation is the direct interaction of neutrino magnetic moment  with external field \cite{Fujikawa1980}. This effect was widely discussed in the 1980s (see, e.g., \cite{Voloshin1986,Akhmedov1988}). In particular, the possible impact of the neutrino spin rotation on the solar neutrino problem was considered.

Another reason for the spin rotation is the interaction of neutrino with matter \cite{Lobanov2001} (a review of the original papers can be found in \cite{Studenikin2004}).
Spin precession takes place even in the case when the matter is at rest and not polarized. If the background medium moves relative to the laboratory reference frame or if it is polarized by some external electromagnetic field, then the neutrino helicity can also change. This effect takes place, since there is a preferred direction different from the direction of the neutrino momentum in these cases.

The motion and the polarization of the background medium were for a long time known to influence the neutrino propagation. The motion of the medium was considered, e.g. in \cite{Pal1989}. In \cite{orsesmo1986_en,Semikoz1987_en,Semikoz1989_en} the concept of the induced magnetic moment of the neutrino was used to describe the background polarization. The value of the induced magnetic moment should be calculated for a particular  medium (e.g., in \cite{Nunokawa1997} it was calculated for the medium composed of electrons only). However, it was not until \cite{Lobanov2001} that the external medium was for the first time considered as a factor, which results in an actual spin precession of the neutrino. As the discovery of the spin rotation in the medium was rather unexpected, for a time the researchers did not pay attention to this effect (see, e.g., \cite{Ochman:2007vn}). Only recently the astrophysical applications of the neutrino spin rotation were studied \cite{Volpe2015,Kartavtsev2015,Dobrynina2016,Vlasenko2014,Ternov2016,Kurashvili2017,Ternov2019}. Although the importance of such effect for the neutrino physics is obvious, its interpretation remains  ambiguous \cite{Nieves2018_1}.

Since the character of the neutrino interaction with electromagnetic field depends on the neutrino flavor, there are correlations between the spin rotation and the flavor oscillations. In the framework of the phenomenological theory these correlations were not studied in a mathematically rigorous way.

In all the important issues neutrino should be considered as a relativistic particle. Hence, a consistent quantum description of the spin rotation may be achieved only with the use of a relativistically covariant equation generalizing the Dirac equation. The interaction of the mass eigenstates with the electromagnetic field is described by the Dirac--Pauli equation, and the interaction with the medium is described by a phenomenological equation, which was obtained in \cite{RT2005_en,Studenikin2005}.
Thus, to describe the neutrino evolution
taking into account both the flavor oscillations and the spin rotation, we also need an explicitly covariant equation for the neutrino wave function.

The paper is organized as follows. In Sec. \ref{model} we derive the equation for neutrino evolution in dense medium following \cite{izvu2016_en}, and generalize it taking into account the direct interaction of neutrino  with the electromagnetic field due to its anomalous moments, including neutrino transition magnetic and electric moments.
In  Sec. \ref{QQappr} we obtain a quasi-classical approximation of this equation and get its solution as a matrix exponential. Using these solutions, we derive a general formula for the probabilities of spin-flavor transitions in the case of constant external conditions.
In  Sec. \ref{2fl} we study two special cases, for which explicit solutions are available. Such solutions exist when the neutrino propagates in electromagnetic field or in moving unpolarized matter. All the results of this section are obtained in the two-flavor model.
In  Sec. \ref{disc} we discuss the main characteristics of the neutrino behavior in the cases considered in the previous section.
In  Sec. \ref{Fin} we summarize the main results of the paper. In Sec. \ref{concl} we discuss some phenomenological consequences of the results obtained.

\section{Wave equation}\label{model}
The equation, which describes the spin rotation of the neutrino mass eigenstates \cite{RT2005_en, Studenikin2005}, takes into account only the neutrino interaction with the medium via neutral currents. Hence, it cannot describe the influence of the medium on the flavor oscillations, a major contribution to which is made by the charged currents. Unfortunately, it is impossible to construct a direct generalization of this equation to describe flavor transitions using the phenomenological approach, because the operator, which transforms the mass eigenstates of neutrino to the flavor states, is not unitary, when it is defined by the mixing matrix only (see e.g. \cite{Hannabus2000}).

To obtain a relativistically covariant equation describing both the neutrino flavor oscillations and its spin rotation, we use a modification of the Standard Model \cite{tmf2017_en,lobanov2019}, where all the fermions with equal electroweak quantum numbers are combined in $SU(3)$-multiplets. That is, each of such multiplets consists of a set of three Dirac fermions. In the framework of this approach, wave functions ${\varPsi}^{(i)}(x)$ describe the fermion multiplet $(i)$ as a whole. These wave functions are $12$-component objects and satisfy the modified Dirac equation
\begin{equation}\label{6_1}
\left( \ii\gamma^\mu\partial_\mu{\mathds{I}}  -{\mathds{M}}^{(i)}\right){\varPsi}^{(i)}(x)=0.
\end{equation}
\noindent In this equation ${\mathds{I}}$ is the $3\times 3$ identity matrix, ${\mathds{M}^{(i)}}$
is a Hermitian mass matrix of the fermion multiplet, which can be represented as follows
\begin{equation}\label{7}
{\mathds{M}}^{(i)}=
\sum\limits_{l=1}^{3}{m_{l}^{(i)}}{\mathds{P}}_{l}^{(i)},
\end{equation}
\noindent where the eigenvalues $m_{l}^{(i)}$ of the mass matrix can be identified with the masses of the multiplet components, and the matrices ${\mathds{P}}_{l}^{(i)}$ are orthogonal projectors on the subspaces with these masses. In Eq. \eqref{6_1} the product of the Dirac matrices and the matrices $\mathds{M}^{(i)}$, $\mathds{I}$ is defined as the tensor product. The transformation properties of the solutions of Eq. \eqref{6_1} are described in detail in \cite{lobanov2019}.

The procedure of quantization in this model is well defined. So, we can obtain the Dyson decomposition, which  enables one to construct the perturbation theory in the interaction picture. As a consequence, we can use the standard diagram technique not only in the tree approximation, but also for the computations of higher-order processes including the radiative corrections. Hence, we can write the equation, which is analogous to the Dirac--Schwinger equation of quantum electrodynamics (see, e.g., \cite{B_SH_en}).

Here we consider the neutrino propagation in matter composed of electrons, protons and neutrons ($e, p, n$). That is, we assume that the density of the neutrino flux is small enough to neglect the effect of neutrino collective oscillations, which were studied, e.g., in \cite{deGouvea:2012hg,Pehlivan:2014zua}. Following \cite{izvu2016_en}, we obtain the equation describing the neutrino interaction with matter.
This interaction, associated with forward elastic scattering of neutrinos by the background fermions, in the framework of quantum field theory may be taken into account if we insert the neutrino mass operator in the modified Dirac equation. In the lowest order of the perturbation theory the analog of the Dirac--Schwinger equation for the neutrino multiplet takes the form
\begin{multline}\label{l161}
\left( \ii\gamma^\mu\partial_\mu{\mathds{I}}  - {\mathds{M}^{(\nu)}}\right)
{\varPsi}(x)
+ \ii \,\frac{g^{2}}{8}\int\!\! d^{4}y \,\gamma^{\mu} (1+\gamma^{5}) S^{(e)}(x,y|{\sf{g}})\gamma^{\nu}(1+\gamma^{5}) D^{W}_{\nu\mu}(y-x)
{\varPsi}(y)  \\
- \ii \frac{g^{2}}{8\cos^{2}\theta_{\mathrm{W}}}\,{\mathds{I}}
\!\!\sum\limits_{i=e,p,n}\int d^{4}y \,\gamma^{\mu}(1+\gamma^{5}) D^{Z}_{\mu\nu}(x-y) \\ \times\Tr\left\{\gamma^{\nu}
\left(T^{(i)}-2Q^{(i)}\sin^{2}
\theta_{\mathrm{W}}+T^{(i)}\gamma^{5}\right)S^{(i)}(y,y|{\sf{g}})\right\}
{\varPsi}(x) = 0.
\end{multline}
\noindent  Here $g$ is the weak interaction constant, $\theta_\mathrm{W}$ is the Weinberg angle, $T^{(i)}$  is the weak isospin projection of the background fermion $(i)$,  $Q^{(i)}$ is the electric charge of the background fermion in the units of the positron charge $e$. The Green functions of free ${W}$ and ${Z}$ bosons are denoted as $D^{W}_{\mu\nu}(x-y)$ and $D^{Z}_{\mu\nu}(x-y)$ respectively, and $S^{(i)}(x,y|{\sf{g}})$ are the Green functions of the fermion multiplets in the real-time formalism \cite{Jackiw1974} (see also  \cite{Levinson1985, Borisov1997_en} and references therein) taking into account the external conditions  $\sf{g}$, i.e. the temperature and the chemical potential of the system.

For relatively low neutrino energies, when  ${\cal E}_{\nu}\ll M_{W}^{2}/{\cal E}_{F}\lesssim M_{W}^{2}/T_{f}$, ${\cal E}_{F} \lesssim T_{f}\ll M_{W}$, where ${\cal E}_{F}, T_{f}$ are the Fermi energy and the temperature of the background fermions (see e.g. \cite{Eminov2016_en}), we can use the Fermi approximation. Then
\begin{equation}\label{ll5}
D^{W}_{\mu\nu}(x-y)\approx \frac{g_{\mu\nu}}{M_{W}^{2}} \delta(x-y),\quad D^{Z}_{\mu\nu}(x-y)\approx \frac{g_{\mu\nu}}{M_{Z}^{2}}\delta(x-y)
\end{equation}
\noindent and Eq. \eqref{l161} takes the form
\begin{multline}\label{l6}
\left( \ii\gamma^\mu\partial_\mu{\mathds{I}}  -{\mathds{M}^{(\nu)}}\right){\varPsi}(x)
+ \ii \frac{G_{\mathrm{F}}}{\sqrt{2}}\Big\{\gamma^{\mu}(1+\gamma^{5}) S^{(e)}(x,x|{\sf{g}})\gamma_{\mu}(1+\gamma^{5})
\\
- \gamma^{\mu}(1+\gamma^{5})\,{\mathds{I}}\!\!\!\sum\limits_{i= e,p,n} \!\!\! \Tr\left\{\gamma_{\mu}\left(T^{(i)}-2Q^{(i)}\sin^{2}
\theta_{\mathrm{W}}+T^{(i)}\gamma^{5}\right)S^{(i)}(x,x|{\sf{g}})\right\}\!\!\Big\}\,
{\varPsi}(x) = 0,
\end{multline}
\noindent  where $G_\mathrm{F}$ is the Fermi constant.

The imaginary parts of the Green functions after the summation over the quantum numbers of the background fermions are reduced to the density matrices, that is $S^{(i)}(x,x|{\sf{g}})\Rightarrow -\ii\varrho^{(i)}(x,x|{\sf{g}})$. The structure of the density matrix for spin $1/2$ fermions is well known from the general considerations \cite{Michel1955}. Since now it is necessary to consider the constituent parts of the medium as the components of the multiplets, for the corresponding density matrices we have
\begin{equation}\label{l7}
\varrho^{(i)}(x,x|{\sf{g}})=\sum\limits_{l=1,2,3}{\mathds
P}^{(i)}_{l}\frac{n^{(i)}_{l}}{4p^{0{(i)}}_{l}}
(\gamma_{\alpha}p^{\alpha{(i)}}_{l}+m_{l}^{(i)})
(1-\gamma^{5}
\gamma_{\alpha}s^{\alpha{(i)}}_{l}),
\end{equation}
\noindent  where $m^{(i)}_{l}$ are the masses of the multiplet components, $n^{(i)}_{l}$ are the number densities of the multiplet components, and $p^{\alpha{(i)}}_{l}, s^{\alpha{(i)}}_{l}$ are the averaged $4$-momentum and $4$-polarization of the multiplet components, respectively.

Equation \eqref{l6} may be presented in a more clear form, if we introduce effective potentials, which are associated with the currents $j^{\alpha{(i)}}_{l}$ and polarizations $\lambda^{\alpha{(i)}}_{l}$ of the background fermions of type $(i)$
\begin{equation}\label{l10}
j^{\alpha{(i)}}_{l}=n^{(i)}_{l}\frac{p^{\alpha{(i)}}_{l}}
{p^{0{(i)}}_{l}}=\{\bar{n}^{(i)}_{l} v^{0{(i)}}_{l},\bar{n}^{(i)}_{l}{\bf{v}}^{{(i)}}_{l}\},
\end{equation}
\begin{equation}\label{l11}
\lambda^{\alpha{(i)}}_{l} =n^{(i)}_{l}\frac{s^{\alpha{(i)}}_{l}}{p^{0{(i)}}_{l}}=
\left\{\bar{n}^{(i)}_{l}
({\bb{\zeta}}^{(i)}_{l}{\bf{v}}^{{(i)}}_{l}), \bar{n}^{(i)}_{l}\left({\bb{\zeta}}^{(i)}_{l} + \frac{
{\bf{v}}^{{(i)}}_{l} ({\bb{\zeta}}^{(i)}_{l}{\bf{v}}^{{(i)}}_{l})}
{1+v^{0{(i)}}_{l}}\right)\right\}.
\end{equation}
\noindent In these formulas
$\bar{n}^{(i)}_{l}$ and ${\bb{\zeta}}^{(i)}_{l}
\;(0\leqslant |{\bb{\zeta}}^{(i)}_{l} |^2 \leqslant 1)$
are the number densities and the average value of the polarization vector of the background
fermions in the reference frame in which the
average momentum of fermions $(i)$ is equal to zero. The $4$-velocity of this reference frame is
denoted as $v^{\alpha{(i)}}_{l}
=\{v^{0{(i)}}_{l},{\bf{v}}^{{(i)}}_{l}\}$ .
The currents $j^{\alpha{(i)}}_{l}$ and polarizations $\lambda^{\alpha{(i)}}_{l}$ of the background fermions characterize the medium as a whole. We introduce effective potentials $f^{\alpha{(e)}}$ and $f^{\alpha (\mathrm N)}$, which generalize the potential used in \cite{Wolfenstein1978}, as follows.
The potential
\begin{equation}\label{l12}
f^{\alpha{(e)}} =\sqrt{2}{G}_{{\mathrm F}}\left({j^{\alpha{(e)}}}
-\lambda^{\alpha{(e)}}\right)
\end{equation}
\noindent determines the neutrino interaction with electrons via the
charged currents, while the potential
\begin{equation}\label{l14}
f^{\alpha (\mathrm N)} =\sqrt{2}{G}_{{\mathrm F}}\sum\limits_{i=e,p,n}
\left({j^{\alpha{(i)}}}
\left(T^{(i)}-2Q^{(i)}\sin^{2}
\theta_{\mathrm{W}}\right)-
{\lambda^{\alpha{(i)}}}T^{(i)}\right)
\end{equation}
\noindent determines the neutrino interaction with all fermions
of the medium via the neutral currents.

Thus, the neutrino evolution equation may be written in the form
\begin{equation}\label{l15}
\bigg(\ii\gamma^{\mu}{\partial}_{\mu}{\mathds I}  - {\mathds{M}}-
\frac{1}{2}
\gamma^{\alpha}f_{\alpha}^{(e)}(1 + \gamma^5){\mathds P}^{(e)} -\frac{1}{2}
\gamma_{\alpha}f^{\alpha({\mathrm N})}(1 + \gamma^5)\,{\mathds I} \!\bigg)\,{\varPsi}(x) = 0,
\end{equation}
\noindent where $\mathds{P}^{(e)}$ is the projector on the state of neutrino with the electron flavor, and ${\mathds{M}}\equiv{\mathds{M}}^{(\nu)}$.
Equation \eqref{l15} describes both flavor oscillations and neutrino spin rotation in dense matter. This equation generalizes the equation, used in \cite{RT2005_en,Studenikin2005} to describe the neutrino spin precession.

Since this equation is obtained by reducing the mass operator of neutrino, the range of energies for which it is applicable is limited from above only. That is, Eq. \eqref{l15} is valid for neutrinos of arbitrary low energies, including the relic ones.

Consider the structure of Eq. \eqref{l15}. As already mentioned, the wave function $\varPsi(x)$ is a $12$-component object.
It is convenient to introduce a block structure of this object, that is to define the object as a set of three Dirac bispinors $\psi_{i}(x)$
\begin{equation}
\varPsi(x) = \left( \begin{matrix} \psi_1 (x) \\ \psi_2 (x) \\ \psi_3 (x) \end{matrix} \right).
\end{equation}
Meanwhile, the $\gamma$-matrices in the evolution equation  act on the components of the Dirac bispinors, and the matrices  $\mathds{M}$, $\mathds{P}^{(e)}$ permute the bispinors $\psi_{i}(x)$ in $\varPsi(x)$.

Similarly to the $\gamma$-matrices in the ordinary Dirac equation, the matrices $\mathds{M}$ and
$\mathds{P}^{(e)}$ acting on the vectors in the flavor space, may be written in different representations, related by unitary transformations. We will use the term ``the mass representation'' for the representation, where the mass matrix is diagonal, i.e. the mass matrix takes the form
\begin{equation}
\mathds{M}_{mass} = \left(  \begin{matrix} m_1 &0 & 0 \\ 0& m_2 & 0 \\ 0 & 0& m_3 \end{matrix} \right).
\end{equation}
\noindent We use the term ``the flavor representation'' for the representation, where the projectors on the flavor states are diagonal. That is, the projectors on the flavor states take the form
\begin{equation}
{\mathds{P}}^{(e)}_{fl}=\left(
          \begin{matrix}
            1&0&0 \\
            0 &0&0 \\
            0&0&0  \\
          \end{matrix}
        \right),\quad
        {\mathds{P}}^{(\mu)}_{fl}=\left(
          \begin{matrix}
            0&0&0 \\
            0 &1&0 \\
            0&0&0  \\
          \end{matrix}
        \right),\quad
        {\mathds{P}}^{(\tau)}_{fl}=\left(
          \begin{matrix}
            0&0&0 \\
            0 &0&0 \\
            0&0&1  \\
          \end{matrix}
        \right).\quad
\end{equation}
\noindent The mass matrices and the projectors in the defined above representations are connected by the Pontecorvo--Maki--Nakagawa--Sakata mixing matrix ${\mathds{U}}$ (see \cite{pdg2018})
\begin{equation}
\mathds{P}^{(l)}_{mass} = {\mathds{U}}^{\dag}\mathds{P}^{(l)}_{fl}{\mathds{U}},\quad \mathds{M}_{fl} = {\mathds{U}}\mathds{M}_{mass}{\mathds{U}}^{\dag}.
\end{equation}

Every solution of Eq. \eqref{l15} corresponds to some neutrino state. We define the mass eigenstates of the neutrino as the states described by the wave functions $\varPsi'_i(x)$ ($i = 1,2,3$), which take the following form in the mass representation at any space-time point
\begin{equation}\label{WF_m}
\varPsi'_1(x) = \left( \begin{matrix} \psi'_1(x) \\ 0 \\ 0 \end{matrix} \right), \qquad
\varPsi'_2(x) = \left( \begin{matrix} 0\\ \psi'_2(x) \\ 0 \end{matrix} \right), \qquad
\varPsi'_3(x) = \left( \begin{matrix} 0 \\ 0\\ \psi'_3(x) \end{matrix} \right).
\end{equation}
\noindent We define the states of neutrino with a particular flavor at a definite space-time point as the states described by the wave functions $\varPsi_i(x)$ ($i = 1,2,3$), which take the following form in the flavor representation at a given point
\begin{equation}\label{WF_f}
\varPsi_1(x) = \left( \begin{matrix} \psi_1(x) \\ 0 \\ 0
\end{matrix} \right), \qquad
\varPsi_2(x) = \left( \begin{matrix} 0\\ \psi_2(x)\\ 0
\end{matrix} \right), \qquad
\varPsi_3(x) = \left( \begin{matrix} 0 \\ 0\\ \psi_3(x)
\end{matrix} \right).
\end{equation}

It should be emphasized that if the potential $f^{\alpha{(e)}}$ (see Eq. \eqref{l12}) is not equal to zero, the mass matrix does not commute with the operator of  Eq. \eqref{l15}. Therefore, in contrast to the vacuum case, for the neutrino interacting with the medium via the charged currents the mass states cannot be properly defined. In other words, the solutions of Eq. \eqref{l15} cannot take the form \eqref{WF_m} at all the space-time points, and so the mass states do not exist.

A further generalization of  neutrino evolution equation \eqref{l15} can be constructed by taking into account the interaction of neutrino with electromagnetic field. As already mentioned, being a massive particle, a Dirac neutrino is characterized by its anomalous magnetic moment. Due to the effect of mixing, neutrinos are created and detected in flavor states, which are different from the mass eigenstates. However, the magnetic moments for flavor neutrinos, i.e. for neutrinos with indefinite mass, are not properly defined. Thus, it is more convenient to define the magnetic moments for the mass eigenstates of neutrinos. Then, in addition to the diagonal magnetic moments  $\mu^{(i)}$, there are transition magnetic and electric moments ${\mu}^{(ij)}, {\varepsilon}^{(ij)}$, which are non-diagonal elements of the matrices of magnetic and electric moments in the mass representation. The values of these moments were obtained in \cite{Shrock1982} (see also \cite{Giunti2015}) in the framework of the Standard Model. In the lowest order of the expansion in the powers of the ratio $M_l^2/M_W^2$ the magnetic and the electric moments take the form
\begin{equation}\label{xl15}
\begin{array}{l}
{\mu}^{(i)}=m_{i}{\mu}_{0}, \\
\displaystyle{\mu}^{(ij)}=- \frac{m_{i}+m_{j}}{2} \frac{\mu_0}{2}\sum\limits_{l=e,\mu,\tau} \mathds{U}^{*}_{l i} \mathds{U}_{l j} \frac{M_l^2}{M_W^2} , \\ \displaystyle
{\varepsilon}^{(ij)}= \;\ii  \frac{m_{i}-m_{j}}{2} \frac{\mu_0}{2} \sum\limits_{l=e,\mu,\tau} \mathds{U}^{*}_{l i} \mathds{U}_{l j} \frac{M_l^2}{M_W^2},
\end{array}
\end{equation}
\noindent where $M_l$ ($l=e, \mu, \tau$) are the masses of the charged leptons, $M_W$ is the mass of the $W$-boson, and $\mu_0$ is defined by the relation
\begin{equation}
\mu_0 = \frac{3 e G_{\mathrm{F}}}{8\sqrt{2}\pi^2}.
\end{equation}

Therefore, to generalize Eq. \eqref{l15} to the case of neutrino interacting with electromagnetic field, we add terms describing the direct interaction of neutrino with the field due to the anomalous magnetic moments and the transition magnetic and electric moments similarly to the Dirac--Pauli equation
\begin{multline}\label{l16}
\bigg(\ii\gamma^{\mu}{\partial}_{\mu}{\mathds I}  - {\mathds{M}}-
\frac{1}{2}
\gamma^{\alpha}f_{\alpha}^{(e)}(1 + \gamma^5){\mathds P}^{(e)} -\frac{1}{2}
\gamma_{\alpha}f^{\alpha({\mathrm N})}(1 + \gamma^5)\,{\mathds I} \\ - \frac{\ii}{2}\mu_{0}F^{\mu \nu
}\sigma _{\mu \nu }{\mathds{M}}- \frac{\ii}{2}F^{\mu \nu
}\sigma _{\mu \nu }{\mathds{M}}_{h}- \frac{\ii}{2}{}^{\star\!\!}F^{\mu\nu}\sigma _{\mu \nu }{\mathds{M}}_{ah}\!\bigg)\,{\varPsi}(x) = 0.
\end{multline}
\noindent Here $F^{\mu\nu}$ is the electromagnetic field tensor,
${}^{\star\!}F^{\mu\nu} =
-\frac{1}{2}e^{\mu\nu\rho\lambda}F_{\rho\lambda}$  is the dual electromagnetic field tensor. The interaction with ${\mu}^{(ij)}, {\varepsilon}^{(ij)}$ is taken into account by introducing the Hermitian  matrices of transition moments ${\mathds{M}}_{h}$ and ${\mathds{M}}_{ah}$. In the mass representation these matrices take the form
\begin{equation}
\displaystyle{\mathds{M}}_{h}\;=\frac{1}{2}\left( \begin{matrix}
0 & (m_1+m_2)  k_{12} & (m_1+m_3) k_{13} \\ (m_2+m_1)k_{21} & 0 & (m_2+m_3) k_{23} \\ (m_3+m_1) k_{31} & (m_3+m_2) k_{32} & 0
\end{matrix}\right),  \\
\end{equation}
\begin{equation}
\displaystyle{\mathds{M}}_{ah}=-\frac{\ii}{2}\left( \begin{matrix}
0 &(m_1-m_2)k_{12} & (m_1-m_3) k_{13} \\ (m_2-m_1) k_{21} & 0 & (m_2-m_3) k_{23} \\ (m_3-m_1) k_{31} & (m_3-m_2) k_{32} & 0
\end{matrix}\right),
\end{equation}
\noindent where
\begin{equation}\label{kij}
k_{ij}=-\frac{\mu_0}{2}\sum\limits_{l=e,\mu,\tau} \mathds{U}^{*}_{l i} \mathds{U}_{l j} \frac{M_l^2}{M_W^2}.
\end{equation}
\noindent The matrices $\mathds{M}_{h}$, $\mathds{M}_{ah}$  in the flavor representation may be obtained with the use of the mixing matrix ${\mathds{U}}$.

Thus, Eq. \eqref{l16} describes neutrino propagation in matter composed of electrons, protons and neutrons in the presence of electromagnetic field. It takes into account both the modification of the flavor oscillations and the spin rotation phenomenon due to the forward elastic scattering by the background fermions and to the interaction with the electromagnetic field.
Eq. \eqref{l16} provides an opportunity to study the correlations between these phenomena in a mathematically rigorous way.

Equation \eqref{l16} was derived for the case when the external conditions are changing rather slowly. When the characteristics of the medium and the electromagnetic field are changing rapidly, one should use the approach described in \cite{Foldy1952} (see also \cite{arlomur}). However, the matrix structure of the equation remains the same.

Note, that even for the solution with the constant external condition there is an important application. If we know the exact solutions of a wave equation, then we are able to calculate the probabilities of different processes of neutrino production using the technique, analogous to the Furry picture in quantum electrodynamics.

\section{Quasi-classical approximation}\label{QQappr}

Using Eq. \eqref{l16} we can study the behavior of neutrinos of arbitrary low energies including the relic neutrinos. However, all the main experimental results in neutrino physics were obtained in the energy range $m_{l}^{2}/{\cal E}_{\nu}^{2}\ll 1$, when the phenomenological approach is also valid.

In this case we can use the quasi-classical approximation to describe the neutrino evolution. Since for the ultra-relativistic particles de Broglie wavelength is small, we can interpret $x^\mu$ not as the coordinates of the event space, but as the coordinates of the neutrino.
If we consider the neutrino multiplet moving with a constant $4$-velocity  $u^\mu$  ($u^2 =1 $), then we can make the substitution  $x^\mu = \tau u^\mu$. It means that the neutrino evolution is characterized by its proper time $\tau$ only. The proper time is related to the path length $L$ as follows
\begin{equation}\label{t02}
\tau = L/|{\bf u}| .
\end{equation}

By analogy with the quasi-classical spin wave functions \cite{Lobanov2006}, we introduce quasi-classical spin-flavor wave functions ${\varPsi}(\tau)$, which describe spin-flavor coherent neutrino states.
The corresponding evolution equation follows from Eq. \eqref{l16}, if we make the substitution
\begin{equation}\label{t2}
 \gamma^{\mu}\partial_{\mu} \Rightarrow \gamma^{\mu}\left(\frac{\partial\tau}{\partial x^{\mu}}\right)\frac{d}{d\tau}={\gamma^{\mu}u_{\mu}}\frac{d}{d\tau}.
\end{equation}
\noindent It should be noted that  substitution \eqref{t2} is possible only when $u^{\mu}=\const$. Since the quasi-classical spin-flavor wave functions are required to satisfy the condition  $\gamma^{\mu}u_{\mu}{\varPsi}(\tau)={\varPsi}(\tau)$, the evolution equation takes the form
\begin{equation}\label{t3}
\left( \ii{\mathds I}\frac{d}{d\tau}  - {\mathcal{F}}\right)\varPsi (\tau)=0,
\end{equation}
\noindent where
\begin{multline}\label{xl17}
{\mathcal{F}}={\mathds{M}}+
\frac{1}{2}(f^{(e)}u){\mathds P}^{(e)}+\frac{1}{2}(f^{({\mathrm N})}u){\mathds I}+ \frac{1}{2}{R}_{e}{\mathds P}^{(e)}\gamma^{5}\gamma^{\sigma}s^{(e)}_{\sigma}
\gamma^{\mu}u_{\mu}+\frac{1}{2}{R}_{{\mathrm N}}{\mathds I}\gamma^{5}
\gamma^{\sigma}s_{\sigma}^{({\mathrm N})}
\gamma^{\mu}u_{\mu}\\
-\mu_{0}\mathds{M}
\gamma^{5}\gamma^{\mu}{\,}^{\star\!\!}F_{\mu\nu}u^{\nu}-\mathds{M}_{h}
\gamma^{5}\gamma^{\mu}{\,}^{\star\!\!}F_{\mu\nu}u^{\nu}+\mathds{M}_{ah}
\gamma^{5}\gamma^{\mu}F_{\mu\nu}u^{\nu}.
\end{multline}

\noindent In Eq. \eqref{xl17} we use the following notations
 \begin{equation}\label{t4}
{R}(f)={\sqrt{(fu)^2 - f^2}}, \quad \displaystyle
s^{\mu}(f)=\frac{u^{\mu}(fu)-f^{\mu}}{\sqrt{(fu)^2-f^2}},
 \end{equation}
 \begin{equation}\label{t5}
 \begin{array}{l} \displaystyle
{R}_e={R}(f^{(e)}), \qquad {R}_\mathrm{N}={R}(f^{(\mathrm{N})}), \qquad
s^\mu_{(e)} = s^\mu(f^{(e)}), \qquad s^\mu_{(\mathrm{N})} = s^\mu (f^{(\mathrm{N})}).
 \end{array}
\end{equation}

In this paper we restrict ourselves to considering external conditions, which do not vary with the space-time point, as it is the first order approximation of a realistic background. It means, we assume that the effective potentials and the electromagnetic field tensor are constant
\begin{equation}
F^{\mu\nu}=\mathrm{const}, \qquad j^{\mu}_f=\mathrm{const}, \qquad \lambda^{\mu}_f=\mathrm{const},
\end{equation}
\noindent For reasons presented in \cite{arlomur}, in this case it is necessary to impose additional constraints on the external conditions.
The strengths of the electric and magnetic
fields and the averaged current and polarization of the medium should obey a self-consistent system of equations. This system of equations consists of the Maxwell equations, the Lorentz equation
\begin{equation}\label{Lor}
\displaystyle
\dot{j}^\mu_f = \frac{e_f}{m_f} F^{\mu}_{\nu} j^\nu_f,
\end{equation}
\noindent and the Bargman--Michel--Telegdi quasi-classical spin evolution equation \cite{BMT}
\begin{equation}\label{BMT}
\displaystyle
\dot{\lambda}^\mu_f = \left( \frac{e_f}{m_f} F^\mu_\nu + 2 \mu_f (g^{\mu\alpha}-v^\mu_f v^\alpha_f) F_{\alpha\nu}\right) \lambda^\nu_f.
\end{equation}
\noindent  In these equations the dot denotes differentiation with respect to the proper time.

From equations \eqref{Lor} and \eqref{BMT} we find the sufficient condition for currents and polarizations to remain constant
\begin{equation}\label{usl}
F_{\mu\nu} j^{\mu}_f = 0, \qquad F_{\mu\nu} \lambda^{\mu}_f = 0.
\end{equation}
Then for the effective potentials defined by Eq. \eqref{l12}, \eqref{l14}, the following relations must be satisfied
\begin{equation}\label{usl2}
F_{\mu\nu}f^{\nu(\mathrm{N})}  = 0, \qquad F_{\mu\nu} f^{\nu(e)} = 0.
\end{equation}
\noindent  The necessary and  sufficient condition for a tensor to have a zero eigenvalue is its second invariant to be equal to zero. Therefore, we consider only the fields, which satisfy the condition
\begin{equation}\label{in}
F_{\mu\nu}{}^{\star\!}F^{\mu\nu}  = 0.
\end{equation}
\noindent Obviously,  condition \eqref{in} is satisfied for the magnetic field, which is the most interesting model for astrophysical applications. In this particular case  relations \eqref{usl} imply that both the velocity of the medium components ${\bf v}$ and the $3$-dimensional vector of polarization ${\bb \zeta}$ are parallel to the vector of magnetic induction.

If the fields and the effective potentials do not depend on the coordinates of the event space, then it is possible to write the solution of Eq. \eqref{l16} using the resolvent  $U(\tau)$
\begin{equation}\label{sf1}
 \varPsi(\tau)=\frac{1}{\sqrt{2 u_0}}\, U(\tau) \varPsi_{0},
\end{equation}
\noindent where the constant $12$-component object $\varPsi_{0}$ has the form
  \begin{equation}\label{sf2}
\varPsi_0=\frac{1}{2}(1-\gamma^{5}\gamma_{\mu}{s}_{0}^{\mu}
) (\gamma_\mu u^\mu+1) \left(\psi^0\otimes e_{j}\right), \quad \bar{\varPsi}_0\varPsi_0= 2.
\end{equation}
\noindent Here $\psi^0$ is a constant bispinor,
 $e_{j}$ is an arbitrary unit vector in the three-dimensional vector space over the field of complex numbers, and ${s}_{0}^{\mu}$ is a $4$-vector of neutrino polarization such that $(u{s}_{0})=0$.

The resolvent in this case takes the form
\begin{equation}\label{0l17}
U(\tau)=e^{-\ii  \mathcal{F}\tau},
\end{equation}
\noindent where the matrix ${\mathcal{F}}$ is defined by Eq. \eqref{xl17}.

Unfortunately, even if  conditions \eqref{usl} are satisfied, it is not possible to find an analytical solution of Eq. \eqref{l16} in the general case, because the problem of calculating the eigenvectors of the matrices in \eqref{l16} results in an algebraic equation of degree no less then six.
The matrix structure of Eq. \eqref{xl17} indicates that even when the effective potentials are independent of the coordinates of the event space in general case the spin and the flavor degrees of freedom cannot be separated. In other words, there are no integrals of motion which characterize neutrino flavor states and neutrino polarization states separately. In general case the neutrino propagates in more complicated spin-flavor states. However, we can calculate the probabilities to detect a neutrino in a state with a definite flavor and a definite projection of the spin on a certain direction. Moreover, a neutrino can be in a mass eigenstate only when there is no interaction with matter via the charged currents and the transition moments are not taken into account. This case is discussed in detail in \cite{arlomur}. Obviously, we can come to the same conclusions, if we describe the neutrino using quantum equation \eqref{l16}.

As was already mentioned, the model of constant fields is the first approximation of a realistic background. It is a rather good approximation, since vector and axial currents for fermions propagating in the constant fields, which satisfy \eqref{in}, in the quasi-classical approximation coincide with those obtained in the quantum description (for more details see \cite{vmu1999}). If the external conditions vary slowly, the adiabatic approximation, based on quasi-classical solutions in the constant fields, gives good results. In \cite{pr2013,Smirnov} an analytical method is developed to describe neutrino propagation in matter in the case, when the dependence of the density on the distance can be considered as several narrow segments, where it changes steeply, separated by wide sloping segments (cliff and valley approximation). This method can also be generalized to the case of interaction with electromagnetic field, if needed. In other cases to consider a realistic environment, one should search for numerical solutions of the quantum equation \eqref{l16}. Note, that in this case the quasi-classical approximation is not valid in general case.

While calculating the transitions probabilities from one state of the neutrino to another, it is convenient to use quasi-classical spin-flavor density matrices introduced similarly to the quasi-classical spin density matrices (see \cite{Lobanov2006})
\begin{equation}\label{rho4}
\rho_{\alpha}(\tau)=\frac{1}{4u^{0}}U(\tau)\big(\gamma^{\mu}u_{\mu}+1\big)\left(1-
  \gamma^{5}\gamma_\mu {s}^{\mu}_{0}
  \right){\mathds P}_{0}^{(\alpha)}\bar{U}(\tau)=\frac{1}{2u^{0}}U(\tau)\big(\gamma^{\mu}u_{\mu}+1\big)
  {\cal{P}}_{0}^{(\alpha)}\bar{U}(\tau).
\end{equation}
\noindent In this formula ${s}^{\mu}_{0}$ defines the initial polarization state of the neutrino, and the projector ${\mathds P}_{0}^{(\alpha)}$ defines its initial flavor state. Thus, ${\cal{P}}_{0}^{(\alpha)}$ is a projector on the initial spin-flavor state of the neutrino. Note, that since in our case the states of the neutrino multiplet are pure states, all the results may be obtained with the help of the wave functions, and using density matrices is convenient though not necessary.
The transition probability  from the state $\alpha$ to the state $\beta$ in the time $\tau$ is determined by the following relation
\begin{equation}\label{ver}
  W_{\alpha \rightarrow\beta}=\Tr\left\{\rho_{\alpha}(\tau)\rho^{\dag}_{\beta}
  (\tau=0)\right\}.
\end{equation}

As it was already mentioned, it is not possible to write an explicit analytical expression for the matrix exponential \eqref{0l17} in the general case. Therefore, we have to use numerical methods to calculate the transition probability. The most effective way to perform such calculations is based on Backer--Campbell--Hausdorff formula. Using this formula, we can write the expression for the transition probability as follows
\begin{equation}\label{ver1}
  W_{\alpha \rightarrow\beta}=\frac{1}{2u^{0}}\Tr \left\{e^{-\ii \tau {\mathcal{F}}}{\cal{P}}^{(\alpha)}_{0}e^{\ii \tau {\mathcal{F}}}{\cal{P}}^{(\beta)}_{0}(\gamma^{\mu}u_{\mu}+1)\gamma^{0}\right\}=
  \frac{1}{2u^{0}}\sum\limits^{\infty}_{n=0}\frac{(-\ii\tau)^{n}}{n!}\Tr \left\{D_{n}{\cal{P}}^{(\beta)}_{0}(\gamma^{\mu}u_{\mu}+1)\gamma^{0}\right\},
\end{equation}
\noindent where
\begin{equation}\label{ver2}
 D_{0}={\cal{P}}^{(\alpha)}_{0},\;  D_{1}=[{\mathcal{F}},{\cal{P}}^{(\alpha)}_{0}],\; D_{2}=[{\mathcal{F}},[{\mathcal{F}},{\cal{P}}^{(\alpha)}_{0}]] ...
\end{equation}

Taking into account the relation between the proper time and the neutrino path length \eqref{t02}, we conclude that the actual expansion parameter in formula \eqref{ver1} is the ratio of the distance between the source and the detector to the flavor oscillation length in vacuum ${L}/{L_{osc}}$, since in the ultra-relativistic limit the value  $\tilde{L}=2 \pi |{\bf{u}}| /{(m_2-m_1)}$ is the standard oscillation length of the phenomenological theory ${L_{osc}}=4 \pi {\cal E}_\nu/{(m_2^{2}-m_1^{2})}$ expressed through another set of quantum numbers.

Expression \eqref{ver1} converges rapidly, since it is a series of exponential type. So, it is convenient to use \eqref{ver1} for calculating the transition probabilities.
This approach provides the opportunity to avoid problems arising in direct calculations of the transition probabilities based on numerical solutions of Eq. \eqref{t3}, since we can avoid working with rapidly oscillating functions.
When calculating  expressions \eqref{ver2}, it is convenient to decompose the flavor projectors using the generators of the $SU(3)$ group, represented as the Gell-Mann matrices, and to use some effective parametrization for the mixing matrix (see, e.g., \cite{Borisov2016}).

\section{Exact solutions}\label{2fl}

Formula \eqref{ver1} can be used to calculate the values of the probabilities in a general case. However, for some models it is possible to represent the matrix exponential in an explicit form.  The study of such models is important for understanding some qualitative properties of neutrino evolution in external conditions.

In particular, the exact solutions may be obtained in two physically interesting cases.
First, an explicit solution of Eq. \eqref{t3} exists when neutrino propagates in unpolarized homogeneous moving medium.
Second, an explicit solution exists, when neutrino interacts with electromagnetic field, if we neglect the transition electric moments. The existence of the spin integrals of motion makes it possible to find the solutions in these cases.
We consider the two-flavor model, in which the probabilities of the transitions take a more simple form than in the realistic three-flavor model.

In the two-flavor model the mass matrix $\mathds{M}$, the matrices of the transition moments $\mathds{M}_{h}, \mathds{M}_{ah}$ and the projector on the electron flavor state $\mathds{P}^{(e)}$ are $2\times 2$ matrices and may be expressed in terms of the Pauli matrices. The corresponding wave function  $\varPsi(\tau)$ is an $8$-component object.
In the mass representation
\begin{equation} \label{M_fl}
\mathds{M} = \frac{1}{2}(\sigma_0(m_1+m_2) - \sigma_3(m_2 - m_1)),
\qquad \mathds{P}^{(e)} =   \frac{1}{2}(\sigma_0 +\sigma_1\sin{2\theta}
+ \sigma_3\cos{2\theta}),
\end{equation}
\begin{equation}\label{M_f2}
\mathds{M}_{h} = \frac{1}{2}(m_1+m_2)\mu_{1}\sigma_1 , \qquad \mathds{M}_{ah} = \frac{1}{2}{(m_1-m_2)}\varepsilon_{1} \sigma_2,
\end{equation}
\noindent
where $\sigma_i, i=1,2,3$ are the Pauli matrices, $\sigma_0$ is the identity $2\times2$ matrix, $\theta$ is the vacuum mixing angle. The value $\mu_1/ \mu_0$ is very small, since it is determined by the ratio of the masses of leptons and the $W$-boson squared (see \eqref{kij}).

To convert the operators,  including the resolvent $U(\tau)$, from the mass representation to the flavor representation, one should use the transformation
\begin{equation}\label{UU}
U(\tau) = \mathds{U} U'(\tau) \mathds{U}^\dag,
\end{equation}
\noindent where $U'(\tau)$ is the resolvent in the mass representation, and $U(\tau)$ is the resolvent in the flavor representation. The Pontecorvo--Maki--Nakagawa-Sakata mixing matrix  $\mathds{U}$ in the two-flavor model is as follows
\begin{equation}\label{U}
\mathds{U} = \left( \begin{matrix} \cos{\theta} & \sin{\theta} \\ - \sin{\theta}
& \cos{\theta} \end{matrix} \right).
\end{equation}

Let us consider the neutrino behavior in the constant homogeneous electromagnetic field. In this case, if we neglect the transition electric moments, for the matrix \eqref{xl17} in the mass representation we have
\begin{equation}\label{xxl17}
{\mathcal{F}}\rightarrow \frac{1}{2}\left\{(\sigma_0(m_1+m_2) - \sigma_3(m_2 - m_1))(1-\mu_{0}
\gamma^{5}\gamma^{\mu}{\,}^{\star\!\!}F_{\mu\nu}u^{\nu})
-\sigma_1 (m_1+m_2) \mu_{1}
\gamma^{5}\gamma^{\mu}{\,}^{\star\!\!}F_{\mu\nu}u^{\nu}\right\}.
\end{equation}
\noindent The spin integral of motion, which defines the projection of the spin on the direction of the magnetic  field in the neutrino rest frame, takes the form
\begin{equation}\label{xxl171}
{\cal S}=-\gamma^{5}\gamma^{\mu}{\,}^{\star\!\!}F_{\mu\nu}u^{\nu}/N,\quad N=\displaystyle \sqrt{\displaystyle u_{\mu}{\,}^{\star\!\!}F^{\mu\alpha}{\,}^{\star\!\!}F_{\alpha\nu}u^{\nu}}.
\end{equation}
\noindent Therefore, the resolvent is given by the relation
\begin{equation}\label{res1}
U'(\tau) = \sum\limits_{\zeta=\pm 1} e^{-\ii\tau T_{\zeta}/2}\big(\cos{(\tau Z_\zeta/2)}  - \ii (X'_\zeta \sigma_1- Y'_\zeta \sigma_3)\sin{(\tau Z_\zeta/2)}\big)\Lambda_\zeta
\end{equation}
\noindent where
\begin{equation}\label{l30x}
\begin{array}{lll}
Y'_{\zeta}&=&\displaystyle\frac{1}{Z_{\zeta}}\Big (\big(m_{2}-m_{1}\big) \big( 1-\zeta\mu_{0}N\big) \Big), \\ [6pt] X'_{\zeta}&=&\displaystyle\frac{1}{Z_{\zeta}}\Big( -\zeta\mu_{1}N\big(m_{2}+m_{1}\big)\Big),\\
Z_{\zeta}&=&\left\{\Big (\big(m_{2}-m_{1}\big) \big( 1-\zeta\mu_{0}N\big) \Big)^{2}+\Big(\big(m_{2}+m_{1}\big)\mu_{1}N\Big)^{2}\right\}^{1/2},\\ [4pt]
T_{\zeta}&=&\displaystyle\big(m_{2}+m_{1}\big) \big( 1-\zeta\mu_{0}N\big).
\end{array}
\end{equation}
\noindent The spin projector has the form
\begin{equation}
\Lambda_\zeta = \frac{1}{2}\left(1 - \zeta {\cal S} \right), \quad [\gamma^\mu u_\mu,\Lambda_\zeta]=0, \quad\zeta=\pm 1.
\end{equation}

The resolvent in the flavor representation may be obtained using  transformation \eqref{UU}
\begin{equation}\label{res11}
U(\tau) = \sum\limits_{\zeta=\pm 1} e^{-\ii\tau T_{\zeta}/2}\big(\cos{(\tau Z_\zeta/2)}  - \ii (X_\zeta \sigma_1- Y_\zeta \sigma_3)\sin{(\tau Z_\zeta/2)}\big)\Lambda_\zeta,
\end{equation}
\noindent where
\begin{equation}\label{l30}
\begin{array}{lll}
Y_{\zeta}&=&\displaystyle\frac{1}{Z_{\zeta}}\Big (\big(m_{2}-m_{1}\big) \big( 1-\zeta\mu_{0}N\big) \cos2\theta+\zeta\mu_{1}N\big(m_{2}+m_{1}\big)\sin 2\theta\Big), \\ [6pt] X_{\zeta}&=&\displaystyle\frac{1}{Z_{\zeta}}\Big (\big(m_{2}-m_{1}\big) \big( 1-\zeta\mu_{0}N\big) \sin 2\theta-\zeta\mu_{1}N\big(m_{2}+m_{1}\big)\cos 2\theta\Big).
\end{array}
\end{equation}
\noindent If we introduce the notations
\begin{equation}
X'_\zeta = \sin{2\theta^{m}_{\zeta}}, \qquad
Y'_\zeta = \cos{2\theta^{m}_{\zeta}},
\end{equation}
\noindent then
\begin{equation}
X_\zeta =\sin{2\theta_{\zeta}}= \sin{2(\theta^{m}_{\zeta}+\theta)}, \qquad
Y_\zeta =\cos{2\theta_{\zeta}}= \cos{2(\theta^{m}_{\zeta}+\theta)}.
\end{equation}
\noindent
That is, $\theta_\zeta$ is an effective mixing angle, which arises when neutrino propagates in electromagnetic field. It is an analog of the famous effective mixing angle in matter \cite{Wolfenstein1978}.

It should be noted that if we do not take into account the transition moments, then $\theta^{m}_{\zeta}=0$. As already mentioned, only in this case we can consider a flavor state of the neutrino as a superposition of the mass eigenstates.

Now we calculate the probabilities of the spin-flavor transitions between different states of the neutrino. For this purpose it is convenient to use the resolvent in the flavor representation, which is given by the relation \eqref{res1}. We consider the transitions between the states with definite flavor. In the flavor representation the projectors on such states take the form
\begin{equation}\label{M_fl2}
\mathds{P}^{(\alpha)}_{0} =   \frac{1}{2}(1 +\xi_{0}\sigma_3), \quad \mathds{P}^{(\beta)}_{0} =   \frac{1}{2}(1 +\xi'_{0}\sigma_3),\quad \xi_{0},\xi'_{0} =\pm 1.
\end{equation}
\noindent To obtain the projectors on the initial and final state with electron flavor one should choose $\xi_0, \xi_0'=1$, otherwise $\xi_0, \xi_0'=-1$. We also assume that in these states neutrino has a definite helicity, i.e.
\begin{equation}\label{t61}
{s}_{0}^{(\alpha)\mu}=\zeta_{0}{s}^{\mu}_{sp},\quad {s}_{0}^{(\beta)\mu}=\zeta'_{0}{s}^{\mu}_{sp},\quad{s}^{\mu}_{sp}=
\{|{\bf u}|,u^{0}{\bf u}/|{\bf u}|\}, \quad \zeta_{0}, \zeta'_{0}=\pm 1,
\end{equation}
\noindent where the values $\zeta_0,\zeta_0'=1$ correspond to the right-handed neutrino in the initial and final state, and $\zeta_0,\zeta_0'=-1$ correspond to the left-handed neutrino. Using  formula \eqref{ver}, we obtain
\begin{multline}\label{l31}
W_{\alpha \rightarrow\beta}=\frac{1+\xi_0\xi'_0}{2}\frac{1+\zeta_0\zeta_0'}
{2}W_1 + \frac{1+\xi_0\xi'_0}{2}\frac{1-\zeta_0\zeta_0'}{2}W_2
+   \frac{1-\xi_0\xi'_0}{2}\frac{1+\zeta_0\zeta_0'}{2}W_3 + \frac{1-\xi_0\xi'_0}{2}\frac{1-\zeta_0\zeta_0'}{2}W_4,
\end{multline}
\noindent where
\begin{equation}\label{l32}
\begin{array}{l}
\displaystyle W_1=\frac{1}{2}\bigg(\frac{1}{2}
(1-\zeta_0(\bar{s} s_{sp}))^2 (1- S_{+1}^2 X_{+1}^2) +\frac{1}{2}(1+\zeta_0(\bar{s} s_{sp}))^2(1-
S_{-1}^2 X_{-1}^2) \\ [8pt]
 \displaystyle {\phantom{W_1=}}+(1- (\bar{s} s_{sp})^2)
(C_{+1} C_{-1}+ S_{+1} S_{-1}
Y_{+1} Y_{-1})\cos({\omega\tau})
 \\ \displaystyle {\phantom{W_1=}}
+\xi_0(1-(\bar{s} s_{sp})^2)(S_{-1} Y_{-1} C_{+1}- C_{-1} S_{+1}Y_{+1})\sin({\omega\tau }) \bigg), \\ \displaystyle
W_2=\frac{1}{2}\bigg( \frac{1}{2}(1-(\bar{s} s_{sp})^2)(2- S_{+1}^2 X_{+1}^2- S_{-1}^2 X_{-1}^2)

   \\ [8pt] \displaystyle {\phantom{W_1=}}- (1-(\bar{s} s_{sp})^2)(C_{+1} C_{-1} + S_{+1} S_{-1} Y_{+1}Y_{-1})\cos({\omega\tau }
  )
   \\ \displaystyle {\phantom{W_1=}}-
\xi_0(1-(\bar{s} s_{sp})^2)(S_{-1} Y_{-1} C_{+1}- C_{-1} S_{+1} Y_{+1})\sin({\omega\tau }) \bigg), \\
\displaystyle
W_3= \frac{1}{2}\bigg( \frac{1}{2}(1-\zeta_0(\bar{s} s_{sp}))^2 S_{+1}^2 X_{+1}^2+ \frac{1}{2}(1+\zeta_0(\bar{s} s_{sp}))^2 S_{-1}^2 X_{-1}^2
\\ [-4pt] \displaystyle {\phantom{W_1=}}+ (1-(\bar{s} s_{sp})^2)S_{+1} S_{-1}X_{+1} X_{-1}\cos({\omega\tau}) \bigg), \\ \displaystyle
W_4= \frac{1}{2}\bigg( \frac{1}{2}(1-(\bar{s} s_{sp})^2)(S_{+1}^2 X_{+1}^2
+ S_{-1}^2 X_{-1}^2)
\\ [-4pt]\displaystyle {\phantom{W_1=}} - (1-(\bar{s} s_{sp})^2)S_{+1} S_{-1}X_{+1} X_{-1}\cos({\omega\tau}) \bigg).
\end{array}
\end{equation}
\noindent  Here
\begin{equation}\label{ff}
\begin{array}{l}
 C_{\pm 1}= \cos\big({\tau Z_{\pm 1}}/2\big), \qquad
S_{\pm 1}=\sin\big({\tau Z_{\pm 1}}/2\big), \;  \\ [6pt]
\omega = \mu_{0}(m_{2}+m_{1})N, \qquad \bar{s}^{\mu}=-{}^{\star\!\!}F^{\mu\nu}u_{\nu}/N.
\end{array}
\end{equation}

The transition probabilities determined by Eq. \eqref{l32} depend on six frequencies. It is quite expected that the probabilities $W_1$ and $W_3$ depend on the initial neutrino polarization $\zeta_0$. What is more interesting, the probabilities $W_1$ and $W_2$ also depend on the initial neutrino flavor $\xi_0$.

If we neglect the transition moment in \eqref{l32} (i.e. set $\mu_{1}=0$), and also assume $(\bar{s} s_{sp})=0$, i.e. consider neutrino moving orthogonally to the magnetic field in the laboratory reference frame, then the expressions for the transition probabilities
$W_{i}^{0}$ are
\begin{equation}\label{l320}
\begin{array}{l}
\displaystyle W_1^{0}=\frac{1}{4}\Big(
2- (S_{+1}^{'2} +
S_{-1}^{'2})\sin^22\theta \\  \displaystyle {\phantom{W_1=}}+
2(C'_{+1} C'_{-1}+ S'_{+1} S'_{-1}
\cos^22\theta)\cos({\omega\tau})  \\
 \displaystyle {\phantom{W_1=}}
+2\xi_0(S'_{-1}  C'_{+1}- C'_{-1} S'_{+1})\cos2\theta\sin({\omega\tau }) \Big), \\ \displaystyle
W_2^{0}=\frac{1}{4}\Big( 2- (S_{+1}^{'2}+ S_{-1}^{'2})\sin^22\theta
  \\ \displaystyle {\phantom{W_1=}}- 2(C'_{+1} C'_{-1} + S'_{+1} S'_{-1}\cos^22\theta)\cos({\omega\tau }
  )  \\ \displaystyle {\phantom{W_1=}}
-2\xi_0(S'_{-1}  C'_{+1}- C'_{-1} S'_{+1} )\cos2\theta\sin({\omega\tau }) \Big), \\
\displaystyle
W_3^{0}= \frac{1}{4}\Big( S_{+1}^{'2}+  S_{-1}^{'2}  + 2S'_{+1} S'_{-1}\cos({\omega\tau}) \Big)\sin^22\theta, \\ [6pt]\displaystyle
W_4^{0}= \frac{1}{4}\Big( S_{+1}^{'2}
+ S_{-1}^{'2}- 2S'_{+1} S'_{-1}\cos({\omega\tau}) \Big)\sin^22\theta,
\end{array}
\end{equation}
\noindent  where
\begin{equation}\label{fff}
 C'_{\pm 1}= \cos\big({\tau Z'_{\pm 1}}/2\big),\;
S'_{\pm 1}=\sin\big({\tau Z'_{\pm 1}}/2\big), \; Z'_{\zeta}= \big(m_{2}-m_{1}\big) \big( 1-\zeta\mu_{0}N\big).
\end{equation}
\noindent If we consider the neutrino with initial electron flavor ($\xi_0=1$), then the formulas  \eqref{l320} coincide with those obtained in \cite{Dvornikov2007,Popov2019}. As was demonstrated in \cite{Dvornikov2007}, the probabilities in this case still depend on six frequencies. However, the dependence on the initial polarization state $\zeta_0$ vanishes.

The values of the transition moments determined by the Standard Model are very small. However, the interaction with the transition moments leads to an interesting effect. The denominators of the functions $Y_{\zeta}=\cos 2\theta_{\zeta}$ and $\quad X_{\zeta}=\sin 2\theta_{\zeta}$, which determine the effective mixing angle, are resonant. As is well known, if the external conditions (in our case, the magnetic induction) vary rather slowly, it can lead to the resonance, which is analogous to the MSW resonance. Note that this is a completely new effect, which was not mentioned in the literature before. The resonance condition $\cos 2\theta_{\zeta} =0$ reduces to the relation $\mu_{0}N=1$, if we neglect the terms proportional to the ratio $\mu_1/\mu_0$.

Let us consider the neutrino propagation in dense unpolarized matter, composed of components moving with the same velocities. In this case the potentials describing the interaction with the medium via charged and neutral currents are proportional
\begin{equation}
f^{\mu(\mathrm{N})} = a f^\mu, \qquad f^{\mu (e)} = f^\mu.
\end{equation}
The coefficient $a$ is determined by the properties of the background fermions
 \begin{equation} \label{a}
 a=\sum\limits_{i=e,p,n}
\frac{n_{(i)}}{n_{(e)}}(T^{(i)}-2Q^{(i)}\sin^{2}
\theta_{\mathrm{W}} ).
\end{equation}

In the two-flavor model  matrix \eqref{xl17} in the mass representation takes the form
\begin{multline}\label{xxxl17}
{\mathcal{F}}\rightarrow\frac{1}{2}\Big\{\sigma_0(m_1+m_2) - \sigma_3(m_2 - m_1)+
\big((fu)+R\gamma^{5}\gamma^{\sigma}s_{\sigma}
\gamma^{\mu}u_{\mu}\big)\big[(\sigma_0 +\sigma_1\sin{2\theta}
+ \sigma_3\cos{2\theta})/2+a\sigma_0\big]\Big\}.
\end{multline}
\noindent Here we use the following notations (see \eqref{t4})
 \begin{equation*}
{R}={\sqrt{(fu)^2 - f^2}}, \quad \displaystyle
s^{\mu}=\frac{u^{\mu}(fu)-f^{\mu}}{\sqrt{(fu)^2-f^2}}.
 \end{equation*}
\noindent The spin integral of motion has the form
\begin{equation}
\tilde{{\cal S}}=\gamma^{5}\gamma^{\sigma}s_{\sigma}.
\end{equation}
\noindent If the medium is at rest, the operator $\tilde{{\cal S}}$ coincides with the helicity operator. The resolvent is written as follows
\begin{multline}\label{sf5}
 \tilde{U}'(\tau) = \frac{1}{2}\sum\limits_{\zeta = \pm 1} \exp{\left\{ - \frac{\ii}{2} \tau \left( (m_2+m_1)
+ \big( (fu) - \zeta R \big)\left(a+\frac{1}{2}\right)
\right)\right\}}\\
\times
\left( \cos(\tau\tilde{Z}_\zeta/{2}) - \ii \sin(\tau \tilde{Z}_\zeta /{2})
\left( \tilde{X}'_\zeta \sigma_1 - \tilde{Y}'_\zeta \sigma_3 \right) \right)(1-\zeta\gamma^{5}\gamma_{\mu}{s}^{\mu}),
 \end{multline}
\noindent where
\begin{equation}\label{sf05}
\begin{array}{l}
\displaystyle \displaystyle  \tilde{Y}'_\zeta = \frac{1}{Z_\zeta} \Big( (m_2-m_1)
 -\big( (fu) -\zeta R \big) \cos{2\theta}/2 \Big), \\[8pt]\displaystyle  \tilde{X}'_\zeta =  \frac{1}{Z_\zeta} \Big(  \big((fu)-\zeta
R\big)\sin{2\theta} /2\Big), \\ [8pt]
\displaystyle \tilde{Z}_\zeta=\sqrt{ \Big(\big( (fu)-\zeta R \big)
/2 -(m_2-m_1)\cos{2\theta}\Big)^{2}+\Big((m_2-m_1)\sin{2\theta}
\Big)^2}.
\end{array}
\end{equation}
\noindent The resolvent in the flavor representation may be obtained from the resolvent in the mass representation using  transformation \eqref{UU}
 \begin{multline}\label{sf07}
 \tilde{U}(\tau) = \frac{1}{2}\sum\limits_{\zeta = \pm 1} \exp{\left\{ - \frac{\ii}{2}\tau
 \left( (m_2+m_1)
+ \big( (fu) - \zeta R \big)\left(a+\frac{1}{2}\right)
\right)\right\}} \\ \times
\left( \cos(\tau\tilde{Z}_\zeta/{2}) - \ii \sin(\tau \tilde{Z}_\zeta /{2})
\left( \tilde{X}_\zeta \sigma_1 - \tilde{Y}_\zeta \sigma_3 \right) \right)(1-\zeta\gamma^{5}\gamma_{\mu}{s}^{\mu}) ,
 \end{multline}
\noindent where
\begin{equation}\label{sf7}
\begin{array}{l}
\displaystyle \tilde{Y}_\zeta = \frac{1}{Z_\zeta} \Big( (m_2-m_1)\cos{2\theta}
 -\big( (fu) -\zeta R \big)/2 \Big),\\[8pt]
\displaystyle \tilde{X}_\zeta =  \frac{1}{Z_\zeta} \Big( (m_2-m_1)\sin{2\theta}
  \Big).
\end{array}
\end{equation}
\noindent If we introduce the notations
\begin{equation}
\tilde{X}'_\zeta = \sin{2\tilde{\theta}^{m}_{\zeta}}, \qquad
\tilde{Y}'_\zeta = \cos{2\tilde{\theta}^{m}_{\zeta}},
\end{equation}
\noindent then
\begin{equation}
\tilde{X}_\zeta =\sin{2\tilde{\theta}_{\zeta}}= \sin{2(\tilde{\theta}^{m}_{\zeta}+\theta)}, \qquad
\tilde{Y}_\zeta =\cos{2\tilde{\theta}_{\zeta}}= \cos{2(\tilde{\theta}^{m}_{\zeta}+\theta)}.
\end{equation}
\noindent

The calculation of the transitions probabilities between the spin-flavor states in this case is quite similar to the calculation for the neutrino in electromagnetic field. Since the structure of the resolvents \eqref{res11} and \eqref{sf07} is identical, the expression for the transition probabilities between the states with definite flavor \eqref{M_fl2} and definite helicity \eqref{t61}, may be obtained from \eqref{l31}, \eqref{l32}, if we make the substitution
\begin{equation}
{X}_\zeta\rightarrow\tilde{X}_\zeta, \quad
{Y}_\zeta\rightarrow\tilde{Y}_\zeta, \quad {Z}_\zeta\rightarrow\tilde{Z}_\zeta, \quad \omega\rightarrow\tilde{\omega}=R(1/2+a),\quad \bar{s}^{\mu}\rightarrow {s}^{\mu}.
\end{equation}
\noindent Thus, the expressions for the transition probabilities coincide with those obtained in \cite{VMU2017_en}. Obviously, in this case the probabilities are also characterized by six frequencies and depend on the initial flavor and polarization state of the neutrino.

Obviously, the denominators of the functions $\tilde{Y}_{\zeta}$ and $\tilde{X}_{\zeta}$, which determine the effective mixing angle in matter $\tilde{Y}_{\zeta}=\cos 2\tilde{\theta}_{\zeta}, \quad \tilde{X}_{\zeta}=\sin 2\tilde{\theta}_{\zeta},$ are resonant, too. For the matter at rest this results in the MSW resonance \cite{MS_en}. If the medium is at rest, then for the left-handed neutrinos  $\tilde{\theta}_{\zeta =-1}$ coincides with the effective mixing angle in matter
\begin{equation}\label{sf71}
\begin{array}{l}
\displaystyle \cos 2\tilde{\theta}_{eff} \approx \frac{ (m_2^{2}-m_1^{2})\cos{2\theta}
 - 2{\cal E}_{\nu}f^{0}}{\sqrt{ \big( 2{\cal E}_{\nu}f^{0}
 -(m_2^{2}-m_1^{2})\cos{2\theta}\big)^{2}+\big((m_2^{2}-m_1^{2})\sin{2\theta}
\big)^2}},\\[12pt]
\displaystyle \sin 2\tilde{\theta}_{eff} \approx  \frac{(m_2^{2}-m_1^{2})\sin{2\theta}}{\sqrt{ \big(  2{\cal E}_{\nu}f^{0}
 -(m_2^{2}-m_1^{2})\cos{2\theta}\big)^{2}+\big((m_2^{2}-m_1^{2})\sin{2\theta}
\big)^2}},
\end{array}
\end{equation}
\noindent since in the ultra-relativistic limit  $|{\bf{u}}| /{(m_2-m_1)}$ coincides with $2{\cal E_\nu}/{(m_2^{2}-m_1^{2})}$.

 Note that for neutrino propagating in matter, as well as for the neutrino interacting with electromagnetic field,  formula \eqref{l32} may be used not only to calculate the probabilities of the transitions between the states with definite helicity, but also to calculate the transition probabilities between the states with arbitrary polarization. For this purpose it is enough to replace $s^{\mu}_{sp}$ in Eq. \eqref{l32} with the desired polarization vector $s^\mu_{0}$. In this case the probabilities may behave in a different way. As already mentioned, the neutrino helicity does not change if the neutrino moves in matter at rest or along the direction of the electromagnetic field. However, if we choose the initial neutrino polarization different from the longitudinal one, even in these cases spin-flip transitions may take place.

\section{Spin rotation}\label{disc}

Equation \eqref{l32} gives the transition probabilities for both neutrino in dense moving matter and in the electromagnetic field. Hence, in these cases the behavior of probabilities is characterized by a number of common properties. Therefore, in this section we use the same notations for the variables $X_{\pm 1}$, $Y_{\pm 1}$, $Z_{\pm 1}$, $\omega$, $s^\mu$ in these cases.

As was already mentioned both in the case of electromagnetic field and in the case of moving medium the probabilities depend on the initial flavor and polarization state of the neutrino, and are characterized by six non-multiple frequencies.
 The frequencies
$Z_{+1}$ and $Z_{-1}$ characterize flavor oscillations of neutrinos with different polarization. Four combinational frequencies $\omega \pm (Z_{+1} \pm Z_{-1})/2$ arise due to correlations between flavor transitions and neutrino spin rotation.
The dependence of the spin-flavor transition probabilities on the distance between the source and the detector has the character of a composite beat.

Due to the properties mentioned, even in the two-flavor model a detailed analysis of the results is rather complicated. Therefore, for clarity, we consider only the spin-flip probability $W_{24}=W_2+W_4$. For neutrinos with initial left-handed polarization the probability $W_{24}$ actually determines the decrease of the total number of neutrinos of all flavors registered experimentally.
Because of the correlations with the flavor transitions, this probability is defined by the expression
\begin{equation}\label{W_per}
  W_{24}=\frac{1}{2}\phantom{.}\mathpzc{A}\phantom{.}(A_1(1 - \cos{\omega_1 \tau}) + A_2 (1- \cos{\omega_2 \tau})
  + A_3 (1 - \cos{\omega_3 \tau}) + A_4 (1 - \cos{\omega_4 \tau})),
\end{equation}
\noindent where the total amplitude of the spin-flip transitions is as follows
\begin{equation}\label{Aaa}
\mathpzc{A}= 1 -(s s_{sp})^2.
\end{equation}
\noindent The probability $W_{24}$  is characterized by four frequencies
\begin{equation}
\begin{array}{l} \displaystyle
\omega_1 = \omega + \frac{Z_{+1}+Z_{-1}}{2}, \qquad
\omega_2 = \omega + \frac{Z_{+1}-Z_{-1}}{2}, \\[10 pt] \displaystyle
\omega_3= \omega - \frac{Z_{+1}-Z_{-1}}{2}, \qquad
\omega_4= \omega - \frac{Z_{+1}+Z_{-1}}{2},
\end{array}
\end{equation}
\noindent and the coefficients corresponding to the oscillating terms are defined by the formulas
\begin{equation}\label{Ai}
\begin{array}{l} \displaystyle
A_1 = \frac{1}{4}(1- Y_{+1} Y_{-1} - X_{+1}X_{-1}+\xi_0 (Y_{+1}-Y_{-1})), \\ [10 pt] \displaystyle
A_2 = \frac{1}{4}(1+ Y_{+1} Y_{-1} + X_{+1}X_{-1}+\xi_0 (Y_{+1}+Y_{-1})), \\[10 pt] \displaystyle
A_3= \frac{1}{4}(1+ Y_{+1} Y_{-1} + X_{+1}X_{-1}-\xi_0 (Y_{+1}+Y_{-1})), \\[10 pt] \displaystyle
A_4= \frac{1}{4}(1- Y_{+1} Y_{-1} - X_{+1}X_{-1}-\xi_0 (Y_{+1}-Y_{-1})), \\[10 pt] \displaystyle
A_1+ A_2 + A_3 + A_4 =1.
\end{array}
\end{equation}

Though the structure of the formulas for the transition probabilities for neutrino interacting with dense medium and with electromagnetic field is similar, these formulas have different physical meaning.
When neutrino propagates in dense matter, the spin-flip probability is limited  above by the total amplitude of the spin-flip transitions
\begin{equation}\label{ampl1}
  \mathpzc{A}=\frac{(v_0^2-1) \sin^2{\vartheta}}{(v_0 u_0 - \sqrt{u_0^2-1}\sqrt{v_0^2-1}\cos{\vartheta})^2 - 1},
\end{equation}
\noindent which depends on the $4$-velocities of the medium $v^\mu$ and the neutrino $u^\mu$. In \eqref{ampl1} $\vartheta$ is the angle between the neutrino velocity and the medium velocity in the laboratory reference frame, $u_0$ and $v_0$ are the Lorentz factors of the neutrino and the medium. Note that the total amplitude does not depend on the number density of the components of the medium.

If the neutrino velocity is greater than the medium velocity, then the total amplitude reaches its maximum value when
\begin{equation}\label{nubol1}
\cos{\vartheta_{max}}=\frac{\sqrt{v_0^2 - 1}/ v_0}{\sqrt{u_0^2 - 1}/u_0},
\end{equation}
\noindent that is when $\cos{\vartheta_{max}}$ is equal to the ratio of the medium velocity to the neutrino velocity. The value of the total amplitude of the spin-flip transitions is equal to
\begin{equation}\label{nubol2}
\mathpzc{A}_{max}=\frac{v_0^2-1}{u_0^2 -1}.
\end{equation}
If the neutrino velocity is less than the medium velocity, then the total amplitude of the spin-flip transitions reaches its maximum value when
\begin{equation}\label{numen1}
\cos{\vartheta_{max}}=\frac{\sqrt{u_0^2 - 1}/u_0}{\sqrt{v_0^2 - 1}/v_0},
\end{equation}
\noindent that is when $\cos{\vartheta_{max}}$ is equal to the ratio of the neutrino velocity to the medium velocity. The value of the total amplitude of the spin-flip transitions is equal to unity
\begin{equation}\label{numen2}
  \mathpzc{A}_{max}=1.
\end{equation}
For the medium at rest the helicity does not change, since in this case $v_0=1$ and the total amplitude of the spin-flip transitions is equal to zero (see \eqref{ampl1}).

\begin{figure}[htbp!]
\begin{minipage}{0.48\linewidth}
\center{\includegraphics[width=\linewidth]{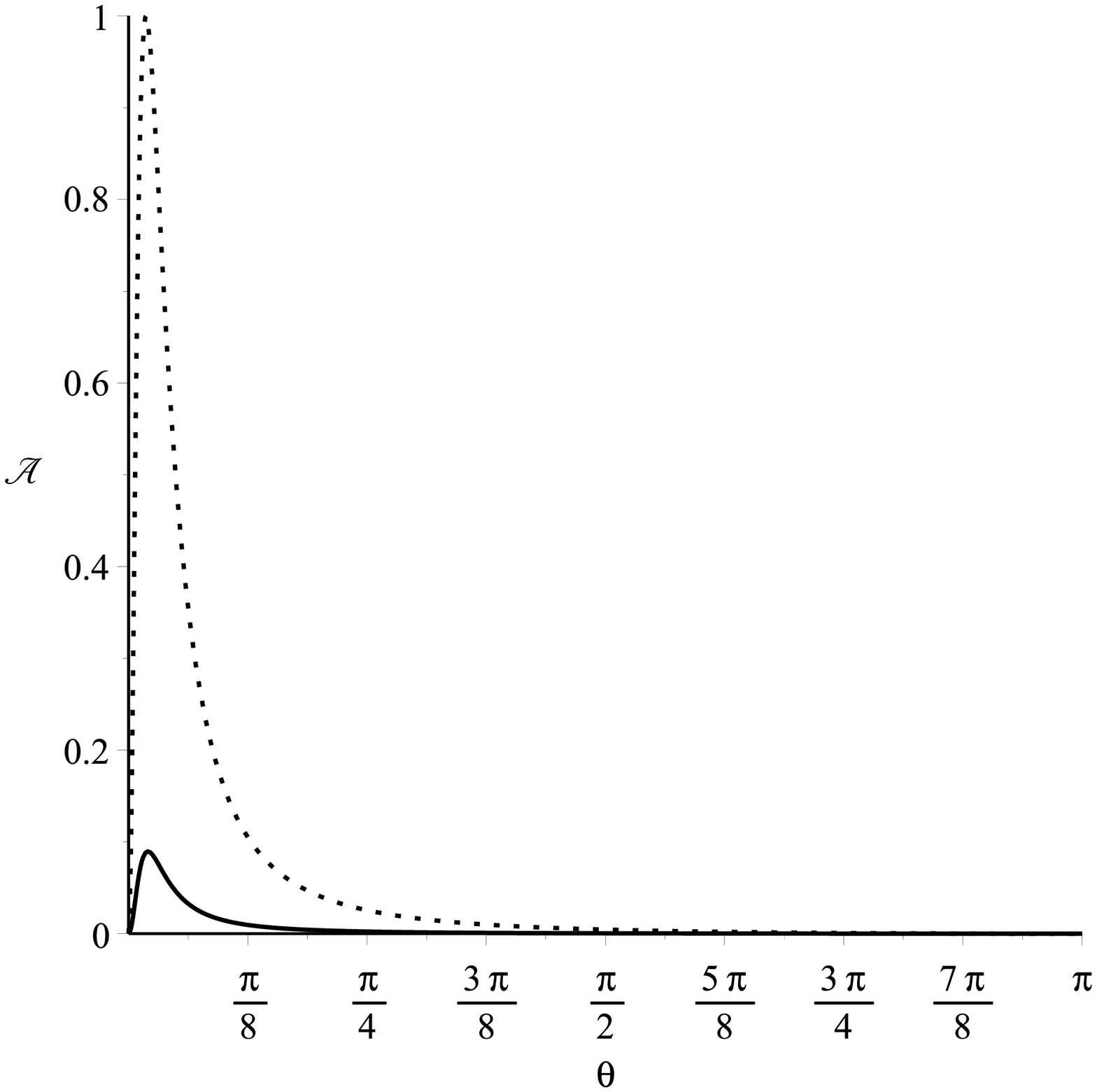}}
\caption{The  total amplitude in matter. The dot line corresponds to $u_0=15, v_0=50$, the solid line corresponds to $u_0=50, v_0=15$.} \label{amplMat}
\end{minipage}
\hfill
\begin{minipage}{0.48\linewidth}
\center{\includegraphics[width=\linewidth]{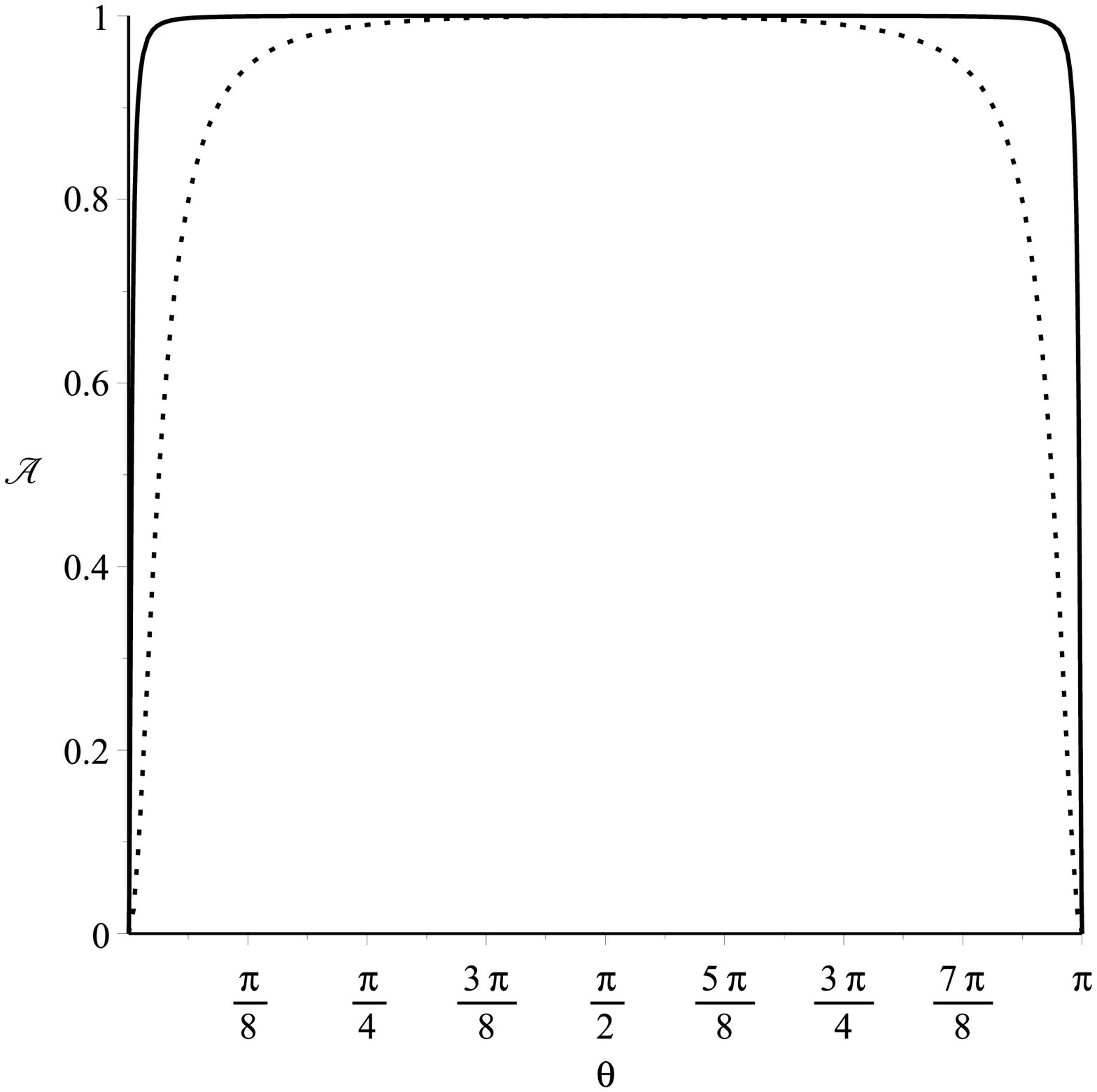} \caption{The  total amplitude in electromagnetic field. The dot line corresponds to $u_0=10$, the solid line corresponds to $u_0=100$.} \label{amplF}}
\end{minipage}
\end{figure}

Equation \eqref{nubol2} implies that if the medium velocity is much less then the neutrino velocity, the probability for neutrino to change its helicity is strongly suppressed. In the medium moving with approximately the same velocity as the neutrino, such probability may reach its maximum value, when the velocities of the neutrino and the medium are almost co-directed (see \eqref{nubol1}-\eqref{numen2}). The dependence of the total amplitude $\mathpzc{A}$ on the angle $\vartheta$ between the directions of motion of the neutrino and the medium is demonstrated in Fig. \ref{amplMat}.

When neutrino propagates in magnetic field, the total amplitude is given by the expression
\begin{equation}\label{ampl2}
\mathpzc{A} = \frac{u_0^2 \sin^2{\vartheta_M}}{u_0^2 - |{\bf u}|^2 \cos^2{\vartheta_M}},
\end{equation}
\noindent where  $\vartheta_M$ is the angle between the direction of neutrino propagation and the magnetic field. That is, the total amplitude depends on the neutrino velocity and the direction of the magnetic-field vector, but does not depend on the value of the magnetic induction.
The total amplitude \eqref{ampl2} is different from unity only when $\vartheta_M\approx 0$ or $\vartheta_M\approx \pi$. Moreover, the angular region, where this difference is essential, contracts when the Lorentz factor of the neutrino $u_0$ increases. The dependence of the total amplitude $\mathpzc{A}$ on the angle $\vartheta_{M}$ is demonstrated in Fig. \ref{amplF}.

Since the transition moment which is equal to $\mu_{1} (m_{1}+m_{2})/2$ is small, far away from the resonance determined by condition $ \mu_{0} N = 1$ the following approximate equalities hold
\begin{equation}
Y_\zeta \approx\cos{2\theta}, \qquad
X_\zeta \approx \sin{2\theta}.
\end{equation}
\noindent Thereby the coefficients corresponding to the oscillating terms are as follows
\begin{equation}
A_1 \approx 0,  \quad \displaystyle
A_2 \approx  \frac{1}{2}(1+\xi_0 \cos{2\theta}),\quad 
A_3 \approx  \frac{1}{2}(1-\xi_0 \cos{2\theta}), \quad \displaystyle
A_4 \approx 0.
\end{equation}
\noindent That is, the beats are effectively characterized by two frequencies $\omega_2$ and $\omega_3$.

When the resonance condition $\mu_0 N=1$ is satisfied, then
\begin{equation}
\begin{array}{l}
Y_{+1}\approx \sin{2\theta}, \qquad X_{+1} \approx  - \cos{2\theta}, \\
Y_{-1} \approx \cos{2\theta}, \qquad X_{-1} \approx \sin{2\theta}.
\end{array}
\end{equation}
\noindent  For the corresponding  oscillating terms we have
\begin{equation}
\begin{array}{l}
\displaystyle A_1 \approx \frac{1}{4}(1+\xi_0 (\sin{2\theta}-\cos{2\theta})), \\ [10 pt] \displaystyle
\displaystyle A_2 \approx \frac{1}{4}(1+\xi_0 (\sin{2\theta}+\cos{2\theta})), \\[10 pt] \displaystyle
\displaystyle A_3 \approx \frac{1}{4}(1-\xi_0 (\sin{2\theta}+\cos{2\theta})), \\[10 pt] \displaystyle
\displaystyle A_4 \approx \frac{1}{4}(1-\xi_0 (\sin{2\theta}-\cos{2\theta})).
\end{array}
\end{equation}
\noindent All these coefficients are non-vanishing. However, in the resonance case only two frequencies are sufficiently different from each other. Therefore, in this case the beats are also effectively characterized by two frequencies only.

\begin{figure}[htbp!]
\begin{minipage}{0.47\linewidth}
\center{\includegraphics[width=\linewidth]{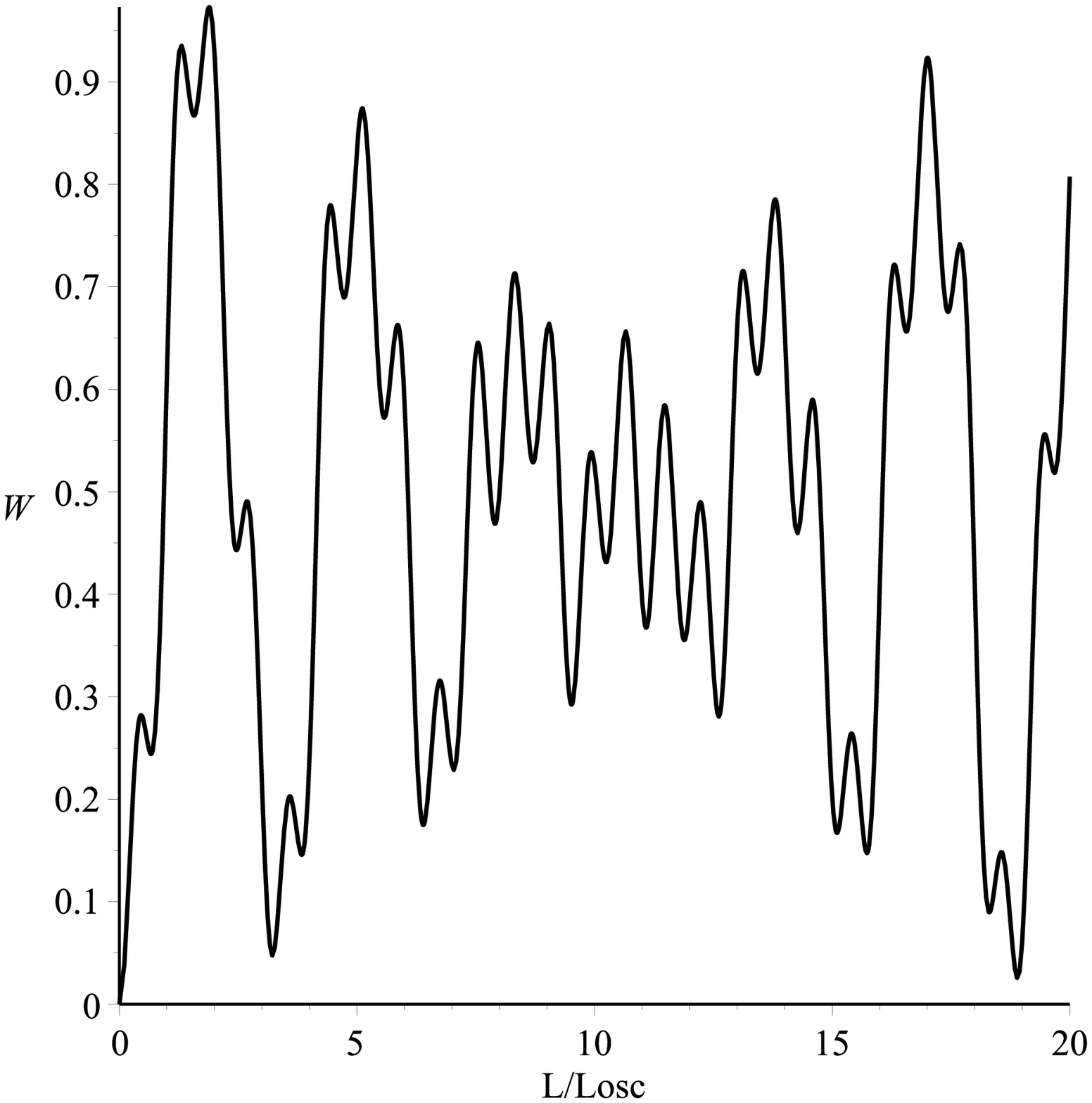}} \caption{Spin-flip probability in matter for $\xi_0= 1$, $u_0=15$, $v_0=50$, $k=10$, $\cos{\vartheta}=\cos{\vartheta_{max}}$} \label{5015k10max_e}
\end{minipage}
\hfill
\begin{minipage}{0.47\linewidth}
\center{\includegraphics[width=\linewidth]{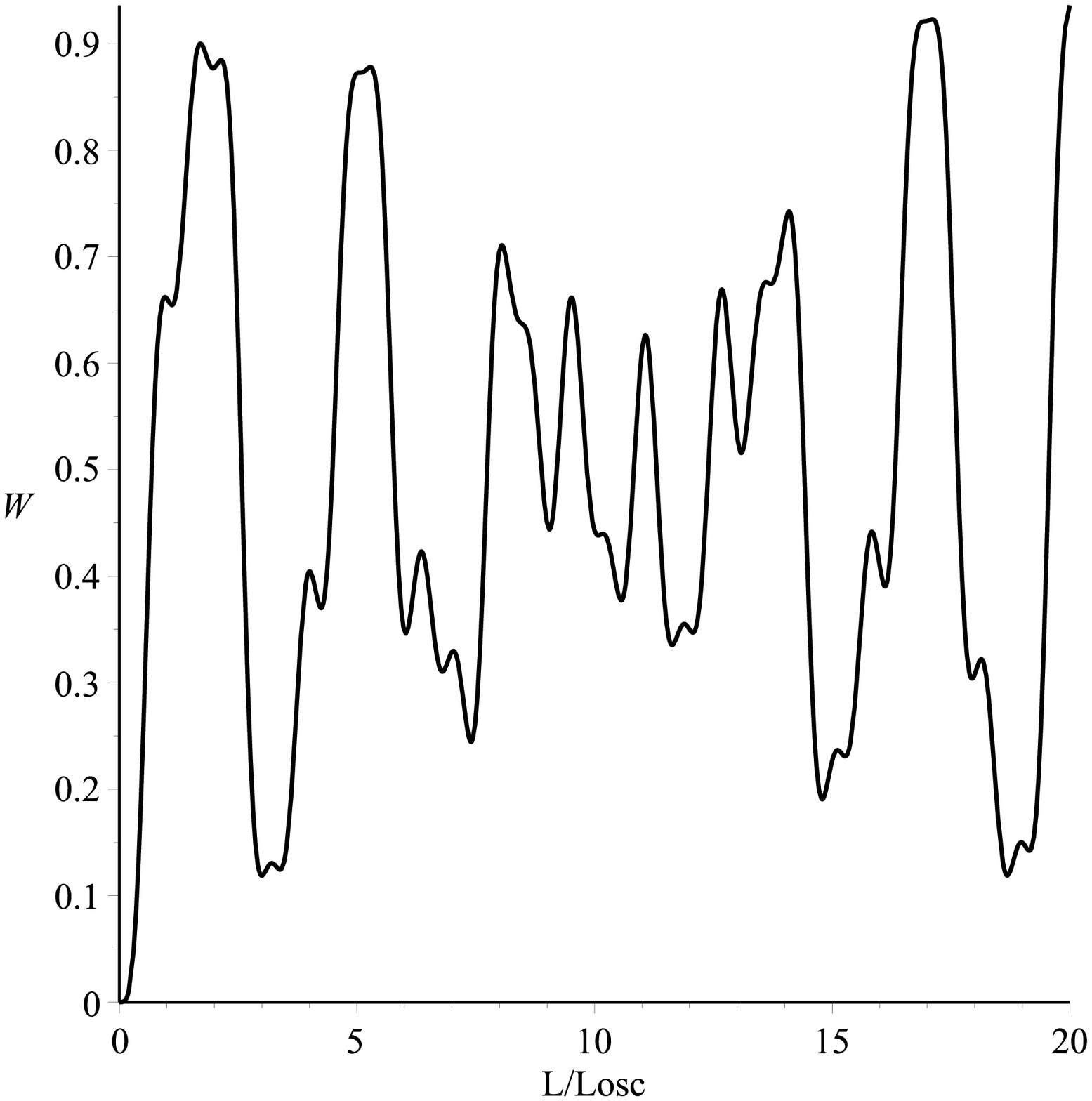} \caption{Spin-flip probability in matter for $\xi_0= -1$, $u_0=15$, $v_0=50$, $k=10$, $\cos{\vartheta}=\cos{\vartheta_{max}}$} \label{5015k10max_mu}}
\end{minipage}
\end{figure}

\begin{figure}[htbp!]
\begin{minipage}{0.47\linewidth}
\center{\includegraphics[width=\linewidth]{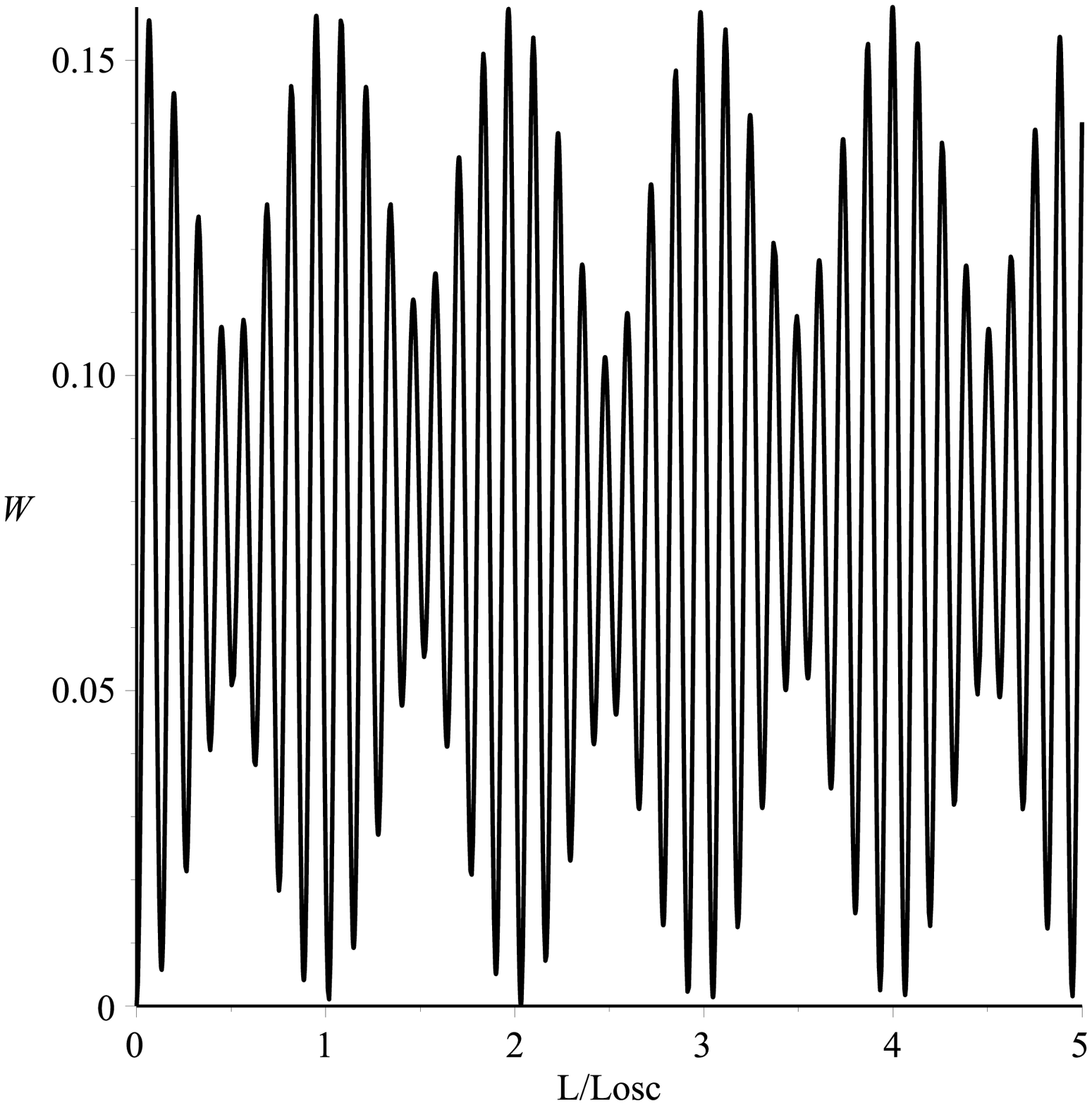}} \caption{Spin-flip probability in matter for $\xi_0=1$, $u_0=15$, $v_0=50$, $k=10$, $\cos{\vartheta}=\,0.95$} \label{5015k10c0.95_e}
\end{minipage}
\hfill
\begin{minipage}{0.47\linewidth}
\center{\includegraphics[width=\linewidth]{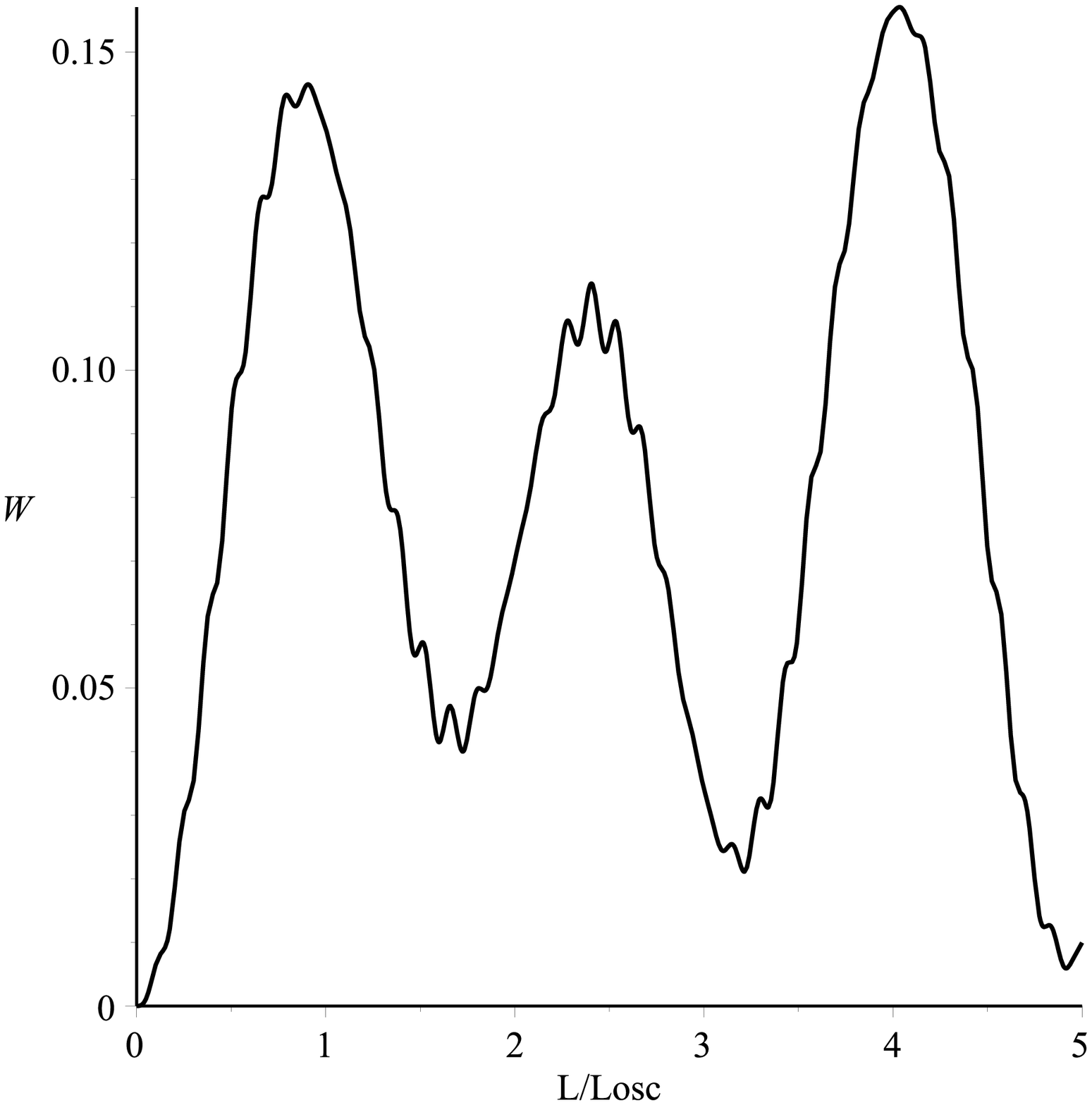} \caption{Spin-flip probability in matter for $\xi_0=-1$, $u_0=15$, $v_0=50$, $k=10$, $\cos{\vartheta}=\,0.95$} \label{5015k10c0.95_mu}}
\end{minipage}
\end{figure}

\begin{figure}[htbp!]
\begin{minipage}{0.47\linewidth}
\center{\includegraphics[width=\linewidth]{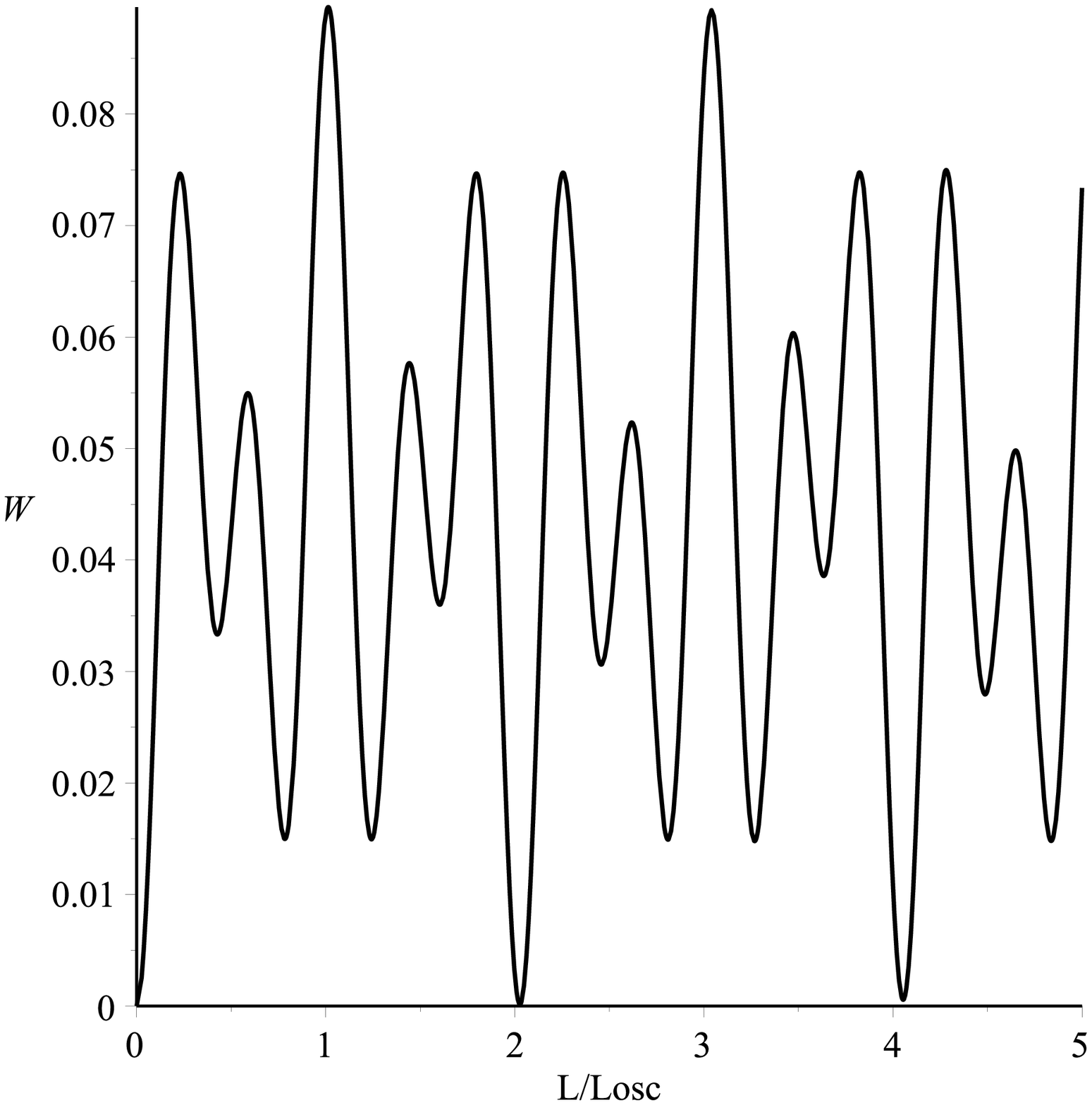}} \caption{Spin-flip probability in matter for $\xi_0= 1$, $u_0=50$, $v_0=15$, $k=10$, $\cos{\vartheta}=\cos{\vartheta_{max}}$} \label{1550k10max_e}
\end{minipage}
\hfill
\begin{minipage}{0.47\linewidth}
\center{\includegraphics[width=\linewidth]{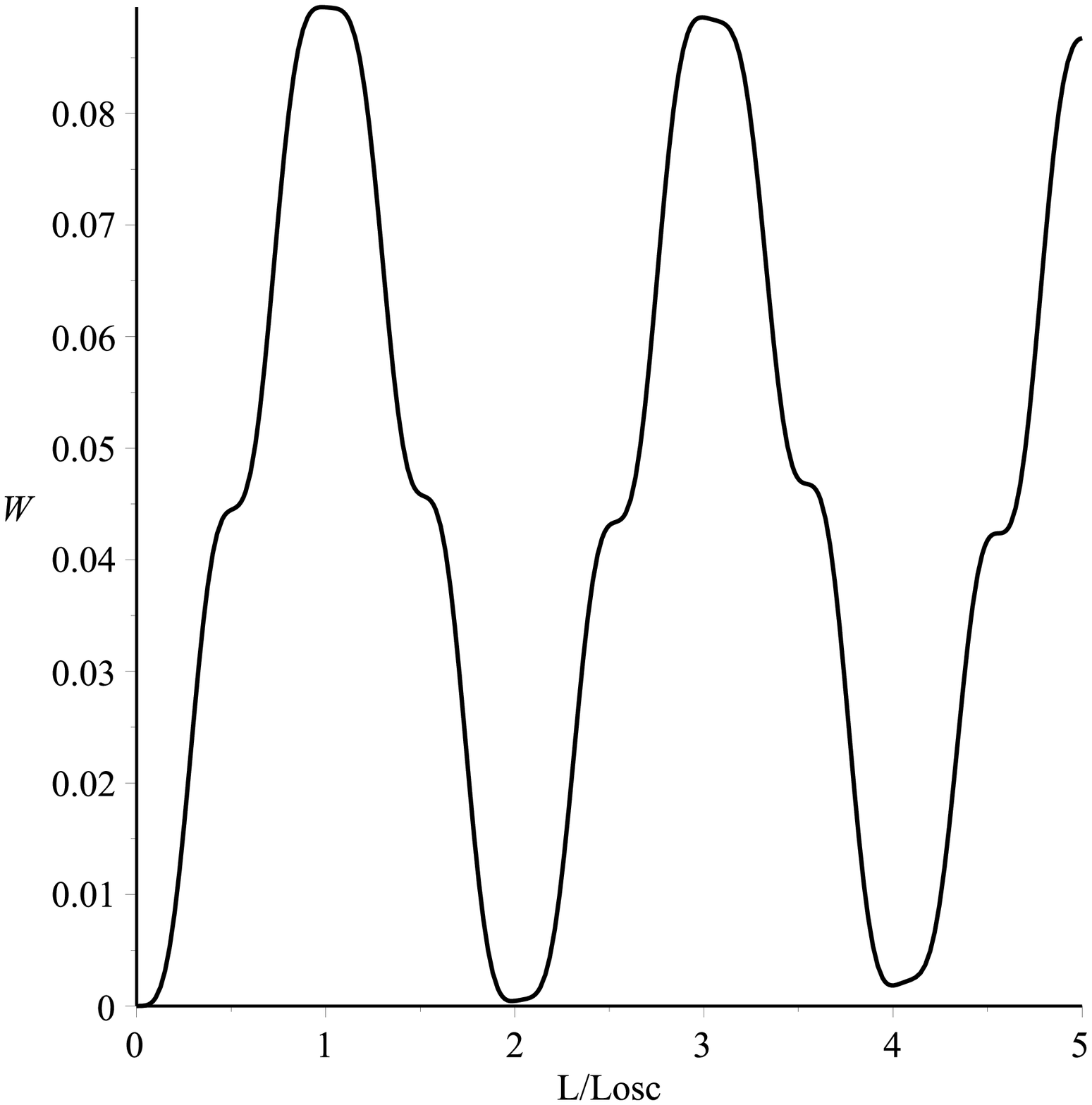} \caption{Spin-flip probability in matter for $\xi_0= -1$, $u_0=50$, $v_0=15$, $k=10$, $\cos{\vartheta}=\cos{\vartheta_{max}}$} \label{1550k10max_mu}}
\end{minipage}
\end{figure}

\begin{figure}[htbp!]
\begin{minipage}{0.47\linewidth}
\center{\includegraphics[width=\linewidth]{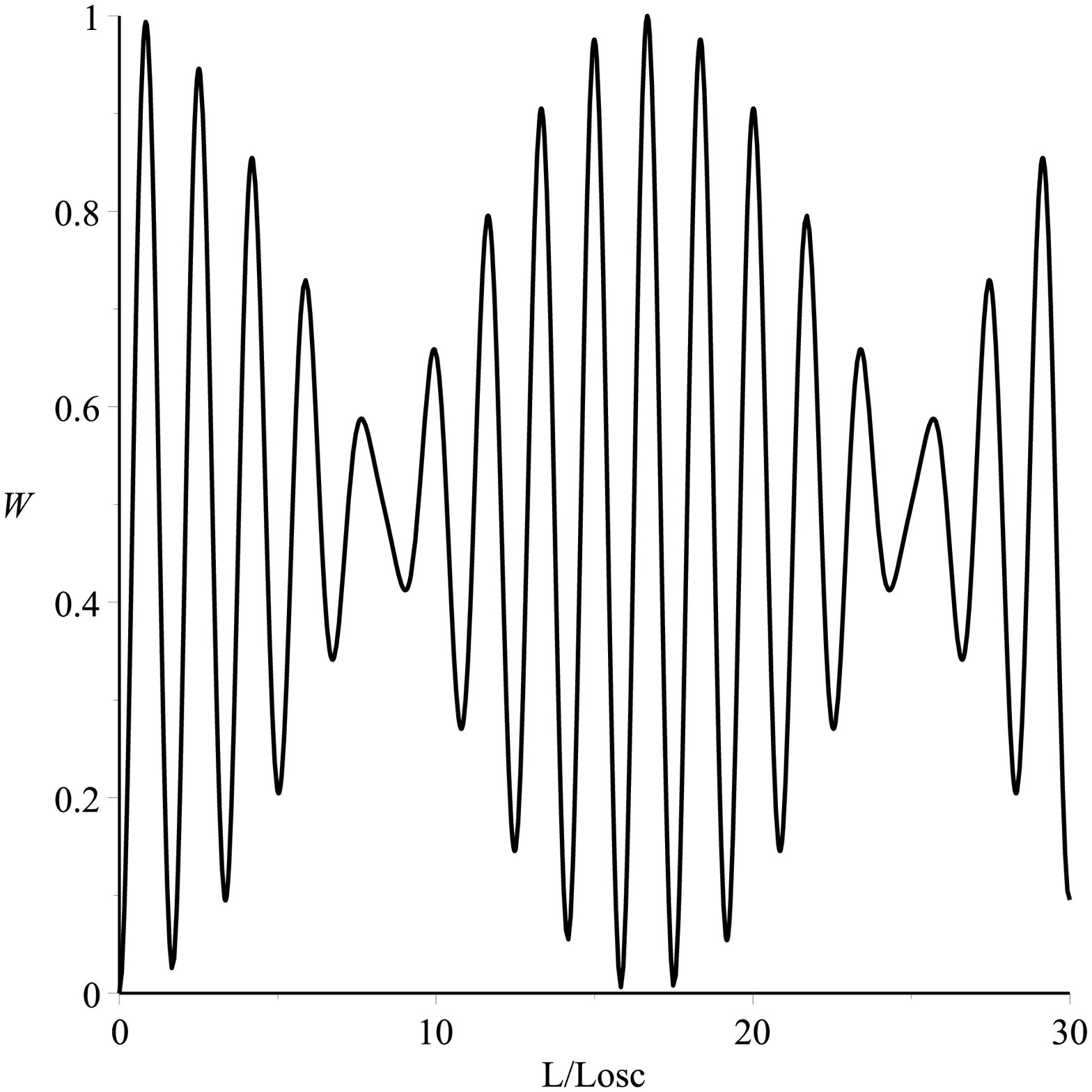}} \caption{Spin-flip probability in magnetic field for $\xi_0= 1$, $\mu_0 N =0.03$, $k_m=20$} \label{magnk003u15_e}
\end{minipage}
\hfill
\begin{minipage}{0.47\linewidth}
\center{\includegraphics[width=\linewidth]{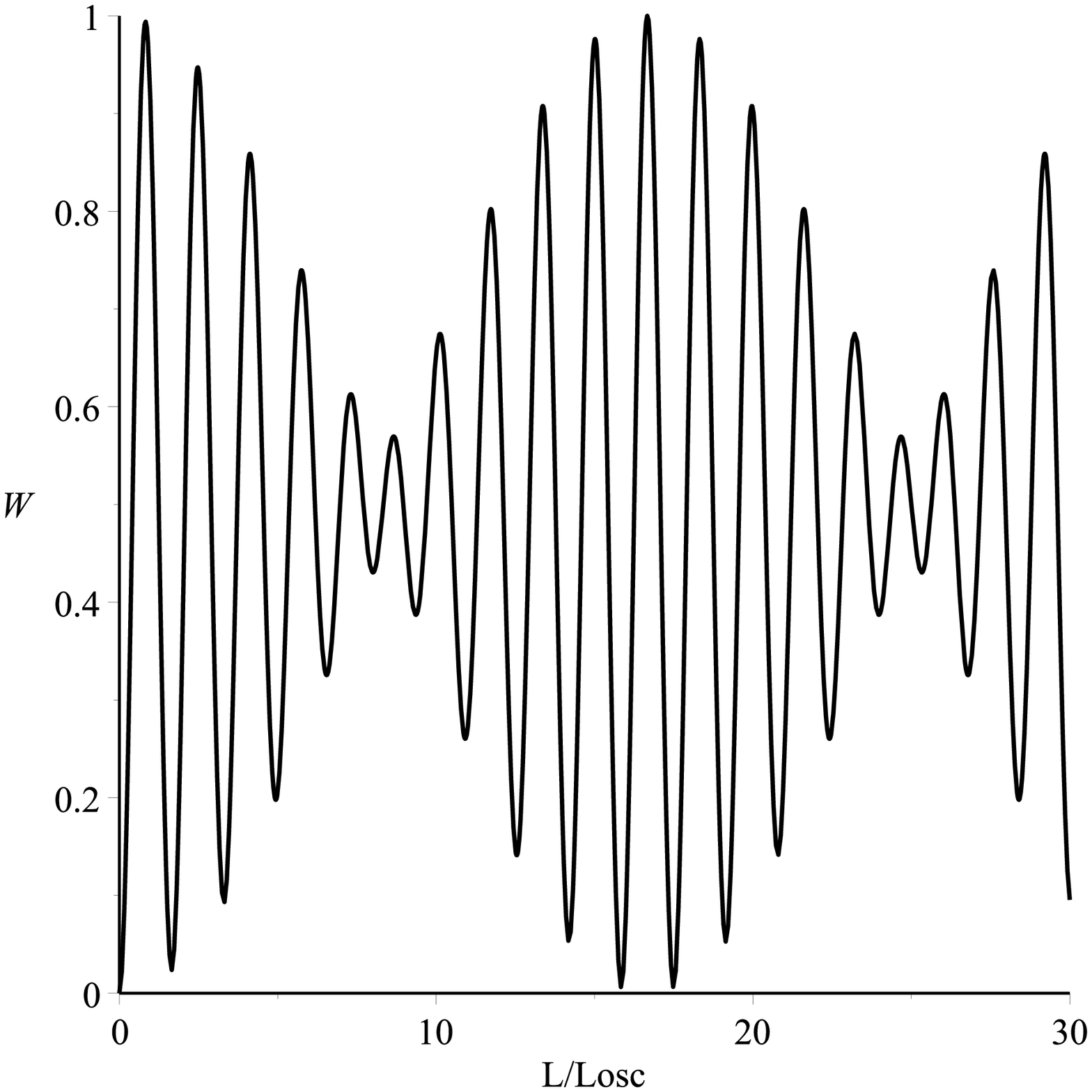} \caption{Spin-flip probability in magnetic field for $\xi_0= -1$, $\mu_0 N =0.03$, $k_m=20$} \label{magnk003u15_mu}}
\end{minipage}
\end{figure}

\begin{figure}[htbp!]
\begin{minipage}{0.47\linewidth}
\center{\includegraphics[width=\linewidth]{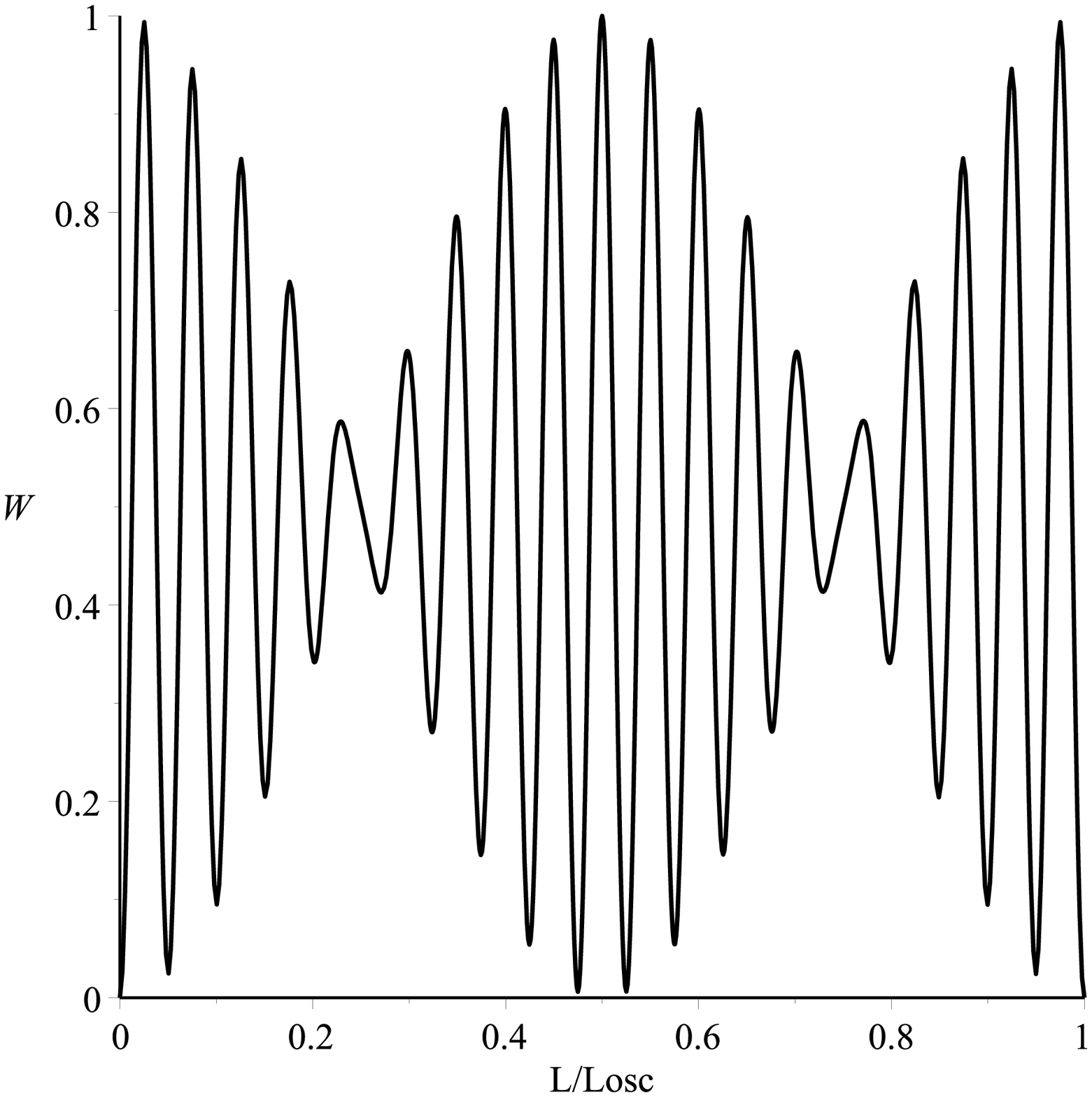}} \caption{Spin-flip probability in magnetic field for $\xi_0=1$, $\mu_0 N =1$, $k_m=20$} \label{magnk1u15_e}
\end{minipage}\hfill
\begin{minipage}{0.47\linewidth}
\center{\includegraphics[width=\linewidth]{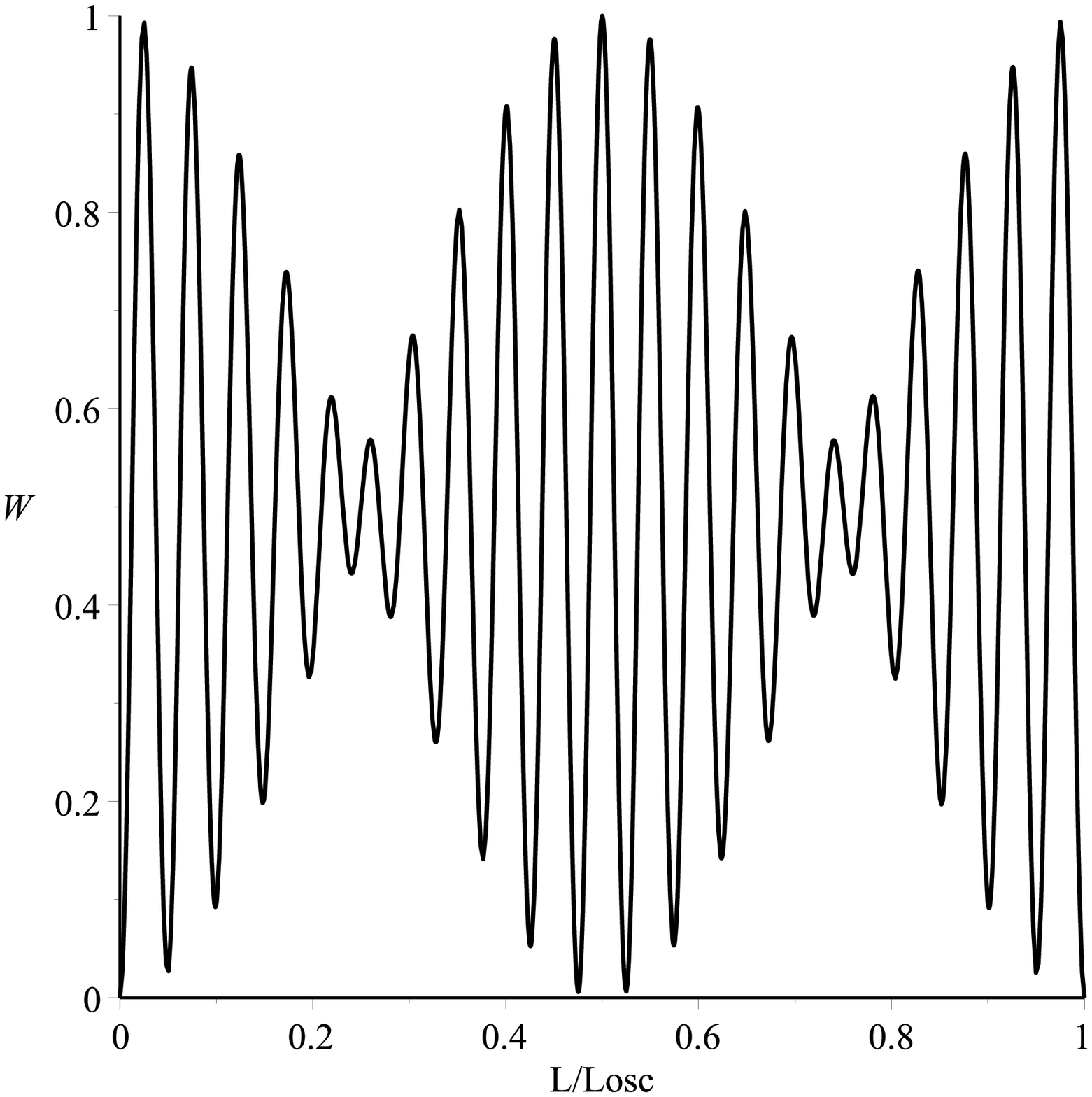} \caption{Spin-flip probability in magnetic field for $\xi_0= -1$, $\mu_0 N =1$, $k_m=20$} \label{magnk1u15_mu}}
\end{minipage}
\end{figure}

To illustrate the main properties of the spin evolution in matter, we study neutrino propagation in medium composed of electrons only. In this case $a=-1/2+2\sin^{2}\theta_{\mathrm{W}}$. The dependence of the spin-flip probability on the distance between the source and the detector $L$ in Figs \ref{5015k10max_e}--\ref{magnk1u15_mu} is given as a function of dimensionless parameter $L/L_{osc}$, where $L_{osc}$ is the flavor oscillation length in vacuum. The behavior of the spin-flip probability depends on the dimensionless parameter
\begin{equation}\label{k}
  k=\frac{\sqrt{2}G_{\mathrm F} n^{(e)}}{|m_2 - m_1|},
\end{equation}
\noindent where $n^{(e)}$ is the electron number density in the laboratory reference frame. The figures are plotted for $\sin^2{\theta}=0.297$ (this corresponds to $\theta_{1 2}$ \cite{pdg2018}).

Figures \ref{5015k10max_e}, \ref{5015k10max_mu}  are given for the case when matter velocity is greater then the neutrino velocity. Here we choose the angle between the neutrino and the matter velocity corresponding to the maximum value of the total amplitude of the spin-flip probability for the chosen values of the Lorentz factors of neutrino and the medium (see \eqref{numen1},\eqref{numen2}). For the chosen values of the velocities $\cos{\vartheta_{max}}\approx 0.998$.
Parameter $k$, which characterizes the medium density, is chosen to be $10$. As can be seen from the figures, the character of the spin oscillations depends significantly on the initial neutrino flavor $\xi_0$.
When the medium density is greater, the dependence on the initial flavor $\xi_0$ becomes more significant.

While Figs. \ref{5015k10max_e}, \ref{5015k10max_mu} are plotted for the angle such that $\cos{\vartheta_{max}}\approx 0.998$,
in Figs. \ref{5015k10c0.95_e}, \ref{5015k10c0.95_mu} the total spin-flip transition probability is plotted for $\cos{\vartheta}=0.95$, i.e.  for a greater value of the angle between the neutrino and the medium velocities. In this case the maximum value of the spin-flip transition probability is much less then unity, and the characteristic frequencies are much greater, then those for $\cos{\vartheta}=\cos{\vartheta_{max}}$.

Figures \ref{1550k10max_e}, \ref{1550k10max_mu} correspond to the case, when neutrino moves faster then the medium. Though these figures are plotted for $\cos{\vartheta_{max}}$, the total amplitude of the spin oscillations is much less then for the cases discussed above.

A specific feature of neutrino propagation in magnetic field is the fact that both for the resonance case $\mu_0 N \approx 1$ and far from the resonance only two frequencies are essential. Similarly to neutrino propagation in dense matter, the spin-flip probability in electromagnetic field depends on the initial neutrino flavor. However, in this case the dependence is less evident.
Figures \ref{magnk003u15_e}, \ref{magnk003u15_mu} correspond to the value $\mu_0 N = 0.03 $, which describes the neutrino behavior far from the resonance. Figures \ref{magnk003u15_e}, \ref{magnk003u15_mu}, \ref{magnk1u15_e}, \ref{magnk1u15_mu} are given for the neutrino propagation orthogonally to the direction of magnetic field. In this case the parameter $N$ and the magnetic induction ${\bf B}$ are connected by the relation $N=|{\bf B}| u_0$. The spin-flip transition probability for neutrino in magnetic field depends on the parameter $k_m=(m_1+m_2)/(m_2-m_1)$. This parameter is not measured experimentally nowadays. Here we choose the value of this parameter $k_m = 20$.

Though the character of the dependence of the probability in Figs. \ref{magnk003u15_e}, \ref{magnk003u15_mu} (far from resonance) and in Figs. \ref{magnk1u15_e}, \ref{magnk1u15_mu} (resonance case) is similar, the typical scale of the spin-flip oscillation in these cases is different.

To compare our results with those obtained in earlier papers, where the spin-flip transitions were studied regardless of the flavor oscillations, in formulas \eqref{l32} we proceed to the limit
\begin{equation}\label{zzzzz}
m_{1}=m_{2}=m.
\end{equation}
\noindent Such assumption is rather unphysical, but it is interesting from the mathematical point of view.

 If neutrino propagates in dense matter and  condition \eqref{zzzzz} is satisfied, then the flavor transitions are absent, which means $W_{3}=W_{4}=0$. For the other probabilities we have
\begin{equation}\label{zl32}
\begin{array}{l}
\displaystyle W_1=\cos^{2}\big(R\tau(a+(1+\xi_{0})/2)/2\big)+
({s} s_{sp})^2\sin^{2}\big(R\tau(a+(1+\xi_{0})/2)/2\big)
  , \\
\displaystyle
W_2=
\big(1-({s} s_{sp})^2\big)\sin^{2}\big(R\tau(a+(1+\xi_{0})/2)/2\big).
\end{array}
\end{equation}
\noindent That means, for neutrino with initial electron flavor both the charged and the neutral currents contribute to the spin-flip transitions. If the neutrino has another initial flavor, then only neutral currents contribute to the spin-flip transitions. In earlier papers the spin behavior was studied for neutrino mass eigenstates. In these papers the charged current interaction was actually taken into account using the operator $1/2 \Tr{(\mathds{P}^{(e)})} \mathds{I} $ instead of the projector $\mathds{P}^{(e)}$.

When neutrino moves in magnetic field, we have
\begin{equation}\label{zzl32}
\begin{array}{l}
\displaystyle W_1=\big(1-\sin^{2}(\mu_{1}mN\tau)\cos^{2}2\theta\big)\big(\cos^{2}(\mu_{0}mN\tau)+
(\bar{s} s_{sp})^2\sin^{2}(\mu_{0}mN\tau)\big), \\[2pt] \displaystyle
W_2=\big(1-\sin^{2}(\mu_{1}mN\tau)\cos^{2}2\theta\big)\big(1-(\bar{s} s_{sp})^2\big)\sin^{2}(\mu_{0}mN\tau), \\
\displaystyle
W_3= \sin^{2}(\mu_{1}mN\tau)\cos^{2}2\theta\big(\cos^{2}(\mu_{0}mN\tau)+
(\bar{s} s_{sp})^2\sin^{2}(\mu_{0}mN\tau)\big), \\ \displaystyle
W_4= \sin^{2}(\mu_{1}mN\tau)\cos^{2}2\theta\big(1-(\bar{s} s_{sp})^2\big)\sin^{2}(\mu_{0}mN\tau).
\end{array}
\end{equation}
\noindent The flavor oscillations in this case are still present, since they are induced by the transition moments. However, the frequency of such oscillations is extremely small. The spin-flip probability does not depend on the initial neutrino
flavor and the spin oscillations are described by the standard formula for rotation of magnetic dipole in
external field.

\section{Summary}\label{Fin}
In conclusion we summarize the main results of the paper.
We obtain the equation for neutrino evolution taking into account neutrino interaction with matter and  with external electromagnetic field. As this equation has no purely spin or purely flavor integrals of motion, we introduce the concept of spin-flavor states of the neutrino, which are described by eigenvectors of a spin-flavor integral of motion.

To describe the evolution of ultra-relativistic particles we consider quasi-classical approximation of this equation. We obtain the formal solution of this equation in the case, when the external conditions do not depend on the coordinates of the event space. Using Backer--Campbell--Hausdorff formula, we develop the general method of calculating the probabilities of the transitions between arbitrary neutrino spin-flavor states.

Then we study neutrino propagation in moving dense matter and in electromagnetic field taking into account the transition magnetic moments, with the use of quasi-classical evolution equation. We find the analytical solutions of the evolution equation and demonstrate, that the expressions for the spin-flavor transition probabilities depend on the initial flavor and polarization state of the neutrino.

We predict resonance behaviour of neutrino in magnetic field due to the transition moments, which was unknown before. Both this resonance and the resonance for neutrino in moving matter, which is a generalization of the famous Mikheev--Smirnov--Wolfenstein resonance, are consequences of the fact that in the general case the neutrino states cannot be described as a superposition of the mass eigenstates, when neutrino propagates in matter and electromagnetic field.

\section{Conclusion}\label{concl}
In the present paper we have studied neutrino flavor oscillation and spin rotation in external fields.
Our approach is based on the modification of the Standard Model put forward in papers \cite{tmf2017_en,lobanov2019}, where neutrinos are Dirac particles. Unfortunately, we cannot construct a mathematically rigorous description of Majorana neutrinos in the same way.
However, from the phenomenological point of view, it seems that to study Majorana neutrinos we only need to change $\gamma^{\mu}f_{\mu}(1+\gamma^{5})/2$ to $\gamma^{\mu}f_{\mu}\gamma^{5}$  and assume $\mu_{0}=0$ in Eq. \eqref{l16}. In this case the formulas for spin-flavor transition probabilities for neutrino in magnetic field will indicate no MSW-like resonance behavior due to transition magnetic moments, which is predicted in this paper for Dirac neutrinos. Hence, the existence of the predicted resonance can become a criterion, which allows us to distinguish between the Dirac and Majorana neutrinos. For this reason here we make some estimates on the magnetic induction and neutrino energies, for which the resonance is expected.

For Dirac neutrino propagating in magnetic field the resonance condition is $\mu_0 N = 1$. That is, if the neutrino propagates orthogonally to the magnetic field, then the resonance is reached when $u_0 (B/B_0)\approx 1.3 \cdot10^{13}$, where $B_0 \approx 4,41 \cdot 10^{13}$ gauss is the Schwinger magnetic field.  That is, $B \approx 5.8\cdot 10^{26} (m_\nu/{\cal E}_\nu)$ gauss, where $m_\nu$ is the average neutrino mass.

Let us consider a magnetar as an example. In a magnetar the magnetic induction can reach the value of $B=10^{16}$ gauss. Let the neutrino mass be $m_\nu = 0.0333$ eV. Then the neutrino energy, which is required for the resonance to take place, is $\mathcal{E}_\nu \approx 1.9$ GeV, which is a very high value.

Hence, a question arises whether the spin-flip effect can be observed at all. For the chosen values of neutrino mass and magnetic induction, the characteristic length of spin oscillations is about $L= \pi /({\mu_\nu B}) \approx  1070$ km, where according to the Standard Model the diagonal magnetic moment of the neutrino is $\mu_\nu = \mu_0 m_\nu \approx 3\cdot 10^{-19} (m_\nu/1 \mathrm{eV}) \mu_B$.  That is much more, than the typical dimensions of magnetars, which is about $R_{mgt}\approx 20-30$ km. Therefore, the spin oscillations are rather unlikely to be observed for a magnetar. Note, that the values of magnetic induction larger than $B=10^{16}$ Gauss are not observed nowadays for any astrophysical objects.

However, this effect can play a significant role in the early Universe, since the values of the fields could be very high. Due to the effect of moving matter on neutrino spin rotation, the spin-flip phenomenon can also be observed for neutrino propagating in galactic jets.

As is well known, there are models of New Physics, which predict greater values of neutrino magnetic moments. Note, that the present experimental limit on the neutrino magnetic moment is $\mu_\nu<2.9\cdot 10^{-11} \mu_B$ \cite{Agostini2017, Beda2013}. For the chosen value of neutrino mass $m_\nu$ that is about $9$ orders of magnitude higher, than the Standard Model theoretical prediction. To take into account such New Physics we only need to choose greater values of neutrino magnetic moments in the expressions for transition probabilities. Therefore, if the New Physics exists, then near magnetars or even near some neutron stars the spin-flip effect may be observed. In this case  the absence of the resonance discussed above will mean that neutrinos are Majorana particles.

\acknowledgments

The authors are grateful to A. V. Borisov,  A. O. Starinets, I. P. Volobuev and V. Ch. Zhukovsky for fruitful discussions. A. V. Chukhnova acknowledges support from the Foundation for the advancement of theoretical physics and mathematics ``BASIS'' (Grant No. 19-2-6-100-1).

\end{document}